\documentclass[12pt]{article}
\pdfoutput=1
\usepackage{a4wide}
\usepackage[centertags]{amsmath}
\usepackage{amssymb}
\usepackage[sort&compress,numbers]{natbib}
\usepackage{ifpdf}
\usepackage{setspace}
\usepackage{amsfonts}
\usepackage{color}
\usepackage{threeparttable}
\usepackage{cancel}
\usepackage{tikz}
\usetikzlibrary{decorations.pathmorphing}

\usepackage{yfonts}

\usetikzlibrary{decorations.markings}
\ifpdf
\usepackage[colorlinks=true,
linkcolor=black,
citecolor=black,
urlcolor=blue,
filecolor=blue,
pdfstartview=FitV,
pdftitle={},
pdfauthor={},
pdfsubject={Hydrodynamics},
pdfkeywords={hydrodynamics, AdS/CFT},
pdfpagemode=None,
bookmarksopen=true
]{hyperref}
\else
\usepackage[hypertex]{hyperref}
\fi
\usepackage{rotating}
\usepackage{rotating}
\makeatletter
\@addtoreset{equation}{section}

\makeatletter
\renewcommand\section{\@startsection {section}{1}{\z@}%
	{-3.5ex \@plus -1ex \@minus -.2ex}
	{2.3ex \@plus.2ex}%
	{\normalfont\large\bfseries}}
\renewcommand\subsection{\@startsection{subsection}{2}{\z@}%
	{-3.25ex\@plus -1ex \@minus -.2ex}%
	{1.5ex \@plus .2ex}%
	{\normalfont\bfseries}}

\def\sec#1{\S \;\ref{#1}}

\title{{Quantum chaos, pole-skipping and hydrodynamics in a holographic system with chiral anomaly}}
\author{Navid Abbasi$^{a}$\footnote{abbasi@lzu.edu.cn},  \ Javad Tabatabaei$^{b}$\footnote{smjty25@gmail.com }\\
\small{\emph{$^{a}$School of Nuclear Science and Technology, Lanzhou University,}}\\
\small{\emph	{ 
222 South Tianshui Road, Lanzhou 730000, China }} \\
\small{\emph{$^{b}$Department of Physics, Sharif University of Technology,}} \\
\small{\emph{P.O. Box 11365-9161, Tehran, Iran}} \\ [1mm]
}

\begin{document}

\setlength{\baselineskip}{16pt}
\begin{titlepage}
\maketitle

\vspace{-36pt}

\begin{abstract}
	It is well-known that  chiral anomaly can be  macroscopically detected through the energy and charge transport, due to the  chiral magnetic effect. 
    On the other hand, in a holographic many body system, the chaotic modes might be only associated with the energy conservation. 
    This suggests that, perhaps, one can detect microscopic anomalies through the diagnosis of quantum chaos in such systems. To investigate this idea, we consider a magnetized brane in AdS space time with a Chern-Simons coupling in the bulk.  By studying the shock wave geometry in this background, we first compute the corresponding butterfly velocities, in the presence of an external magnetic field $B$, in $\mu\ll T$ and $ B \ll T^2$ limit. We find that the butterfly propagation in the direction of $B$ has a different velocity than in the opposite direction; the difference is $\Delta v_{\text{B
	}}=(\log(4)-1)\Delta v_{sound}$ with $\Delta v_{sound}$ being the difference between the velocity of two sound modes propagating in the system.  The splitting of butterfly velocities confirms the idea that chiral anomaly can be macroscopically manifested via quantum chaos. We then show that the pole-skipping points of energy density Green's function of the boundary theory coincide precisely with the chaos points. This might be regarded as the hydrodynamic origin of quantum chaos in an anomalous system. 
Additionally, 
 by studying the near horizon dynamics of a scalar field on the above background, we find the spectrum of pole-skipping points associated with the two-point function of dual boundary operator. We find that the sum of wavenumbers corresponding to pole-skipping points at a specific Matsubara frequency is a universal quantity, which is independent of the scaling dimension of the dual boundary operator. We then show that this quantity follows from a closed formula and can be regarded as another macroscopic manifestation of the chiral anomaly.

  \end{abstract}
\thispagestyle{empty}
\setcounter{page}{0}
\end{titlepage}

\renewcommand{\baselinestretch}{1}  
\tableofcontents
\renewcommand{\baselinestretch}{1.2}  
\section{Introduction}

It has been observed that for a large class of many-body systems \cite{LArkin, Shenker:2013pqb,Roberts:2014isa,Shenker:2014cwa,Maldacena:2015waa,kitaev:2014bwb,Kitaev:2017awl,Polchinski:2015cea,Polchinski:2016xgd,Jensen:2016pah,Maldacena:2016hyu,Gu:2016oyy,Davison:2016ngz,Kukuljan:2017xag,Ling:2016ibq,Blake:2017qgd}, 
\begin{equation} \label{OTOC}
\langle \,V(t,\vec{x})W(0,0)V(t,\vec{x})W(0,0)\,\rangle_{\beta}=1-\epsilon \,e^{-i \omega t+ i k |\vec{x}|}+\cdots
\end{equation}
where $V$ and $W$ are generic few-body operators, $\beta=1/T$ and $\epsilon \sim 1/\mathcal{N}$
with $\mathcal{N}$ being the number of degrees of freedom. The chaotic behavior of the system is represented by the purely imaginary values of the $\omega$ and $k$: 
\begin{equation} \label{chaos_point}
\omega= i \lambda,\,\,\,\,\,\,\,\,\,\,\,\,\,\,\,k=i \frac{\lambda}{v_{B}}
\end{equation}
Here $\lambda$ is the quantum Lyapunov exponent and $v_{B}$ is the butterfly velocity, i.e. the velocity at which the information propagates in the space.
The exponential growth of the out-of-time-order-correlator (OTOC) \eqref{OTOC} is known as the manifestation of the quantum chaos in many-body systems.

Holographically, butterfly velocity can be computed via studying an eternal black hole as being dual to a thermofield double state  in a CFT \cite{Maldacena:2001kr}. Let us recall that the chaotic behavior \eqref{OTOC} is related to sensitivity of dynamics to initial conditions. Correspondingly, injecting an small amount of energy into the left side of the eternal black hole, in the past, will have exponentially blue-shifted energy near the horizon. Therefore, the backreaction on the geometry must  be included \cite{Shenker:2013pqb}. The resultant geometry can be described as a Dray-'t Hooft shock wave \cite{Dry}. For a localized shock in $U-V$ coordinates, with $T_{UU}\sim \delta (U)$, the exponential growth of the discontinuity in the $V$ coordinate, when passing through the shock, is then the holographic dual of \eqref{OTOC}. Thus the problem of finding butterfly velocity reduces to finding the shift in the $V$ coordinate via solving Einstein equations in the bulk.

Based on the above holographic picture, in the first part of the paper, we compute the butterfly velocity in a system with chiral anomaly. Quantum chaos in anomalous systems has not been studied in the literature so far. However, our motivation originates from the fact that in such systems, some kinds of hydrodynamic transport are related to the microscopic triangle anomalies \cite{Son:2009tf}. For example in the presence of an external magnetic field, the energy flux encodes some information about the chiral anomaly due to the chiral magnetic effect \cite{Landsteiner:2016led}.\footnote{A possible relevant experimental evidence has  been reported in \cite{Abelev:2009ac}.}
We also know that, in a holographic system, the chaotic modes may be associated with the energy conservation. This suggests that, perhaps, one can detect chiral anomaly via quantum chaos in the holographic systems.

 The holographic dual of a four dimensional system in the presence of an external magnetic field is a magnetized brane in AdS5 space time \cite{DHoker:2009ixq}. In order to make the   boundary system anomalous, it is convenient to couple the magnetized brane to a Chern-Simons term in the bulk. The gauge invariance in the bulk then leads to existence of chiral anomaly on the boundary, through the inflow mechanism \cite{DHoker:2009ixq}. Thus in order to study a thermal state in the boundary chiral system, we need a black brane solution in the Einstein-Maxwell-Chern-Simons theory. Such solution has been already found analytically in the regime of small magnetic field $B$, in \cite{DHoker:2009ixq}. However, the metric functions in \cite{DHoker:2009ixq} are given by some un-evaluated integrals. In order to work with them, we adopt  the double expansion over $B/T^2$ and $\mu/T$, introduced in \cite{Abbasi:2018qzw}, to evaluate the metric functions. Consequently, whatever we find in this paper will be analytic and will be given in a double expansion over these two dimensionless parameters.

Using the shock wave picture, in the first part of the paper, we compute the butterfly velocity in the above-mentioned background, for two cases;
 firstly in the case where information propagates parallel to the magnetic field,\footnote{This corresponds to a situation   in \eqref{OTOC}  where the line connecting the insertion points of the operators $V$ and $W$ in the space is parallel to the magnetic field.} and secondly in the case where it propagates
perpendicular to the magnetic field. In agreement  with our earlier motivation, we show that in the parallel case, the microscopic anomaly is really detected via the butterfly velocities.
While in a non-chiral system for both $x<0$ and $x>0$ in \eqref{OTOC} one obtains the same butterfly velocity \cite{Shenker:2013pqb,Blake:2017qgd}, for a chiral system in the presence of magnetic field parallel to $\vec{x}$, however,  the butterfly velocities in these two ranges have different magnitudes. Their difference, which we refer to as $\Delta v_{B}$, is proportional with the chiral anomaly coefficient times $(\frac{\mu}{T})^2(\frac{B}{T^2})$. That $\Delta v_B$ depends on the anomaly coefficient shows that it could be a new macroscopic measure for detecting the microscopic chiral anomaly. 

In another direction, by computing the spectrum of the hydrodynamic modes in the same bulk background, we find that the magnitude of velocity of sound modes, going parallel and anti-parallel to the magnetic field, differ from each other; to leading order in the double expansion over $\mu/T$ and $B/T^2$, their difference, namely $\Delta v_{sound}$, is related to  $\Delta v_{B}$ as $\Delta v_{\text{B
}}=(\log(4)-1)\Delta v_{sound}$.
On the other hand, any splitting between velocity of sound waves is caused due to the chiral magnetic effect, both in energy and charge transports. One then \textit{may} conclude that the splitting between butterfly velocities has the same origin. Although, to confirm this idea it is needed to show that the mentioned relation between $\Delta v_{B}$ and $\Delta v_{sound}$ continues to hold beyond the perturbative limit on $B$ and $\mu$.\footnote{We thank the anonymous referee to pointing this out to us.}

It has been recently shown that in addition to \eqref{OTOC}, quantum chaos has  a sharp manifestation in the two-point functions of energy density and energy flux. This manifestation was firstly observed in the numerical spectrum of quasi normal modes in a holographic system \cite{Grozdanov:2017ajz}. In the mentioned paper, it has been shown that the values of frequency and momentum given in  \eqref{chaos_point} follow from  a dispersion relation of the hydrodynamic sound mode.  Additionally, it has been shown that just at this point, the residue of retarded two-point function of the energy density vanishes. This observation actually established a direct link between hydrodynamics and  the butterfly effect, for the first time. This so-called "pole-skipping" phenomenon was then derived as a general prediction of effective field theory in \cite{Blake:2017ris}. In the mentioned paper, based on previously developed effective field theory  of dissipative fluids \cite{Crossley:2015evo,Glorioso:2018wxw}\footnote{The effective theory of dissipative fluid  has been also studied in \cite{Haehl:2015pja,Jensen:2017kzi}, differently. Before these references, however, deriving dissipative hydrodynamics from effective action on a Schwinger-Keldysh CTP contour, was originally initiated by \cite{Grozdanov:2013dba}. }, a new quantum theory of hydrodynamics was formulated. This quantum theory is valid not only to all orders in derivative, but also at finite $\hbar$. The latter emphasizes that the theory of quantum hydrodynamics has been constructed for studying those systems for which the Lyapunov exponent is of the order of $\lambda\sim \frac{1}{\hbar}$.

The main idea of  \cite{Blake:2017ris} is that the scrambling of a few body operator can be described as the following: there exists an effective chaos mode which propagates and builds up an exponentially growing hydrodynamic cloud, around the operators.  The exponential growth basically arises from a shift symmetry. 
This shift symmetry  leads to chaotic behavior in out-of-time-order correlators (eq. \eqref{OTOC}), and simultaneously
shields the correlation functions of energy density and energy flux from exponential growth. The latter happens because at exactly the chaos point \eqref{chaos_point}, the residue of theses correlation functions turns out to vanish; this is nothing but the pole-skipping phenomenon. The pole-skipping has been explicitly shown to happen in 2-dim CFT at large central charge \cite{Haehl:2018izb} and recently in higher dimensional CFTs \cite{Haehl:2019eae}. 

The effective field theory approach of \cite{Blake:2017ris} associates the chaotic mode to the energy conservation. Such feature may be universal among the systems which are or are close to being maximally chaotic. In \cite{Blake:2018leo}, some further strong support for the pole-skipping phenomenon in such systems was found. By studying the linearized Einstein equations around  a static black hole geometry, it has been shown that the pole-skipping is universal for general systems at finite temperature, dual to Einstein gravity coupled to matter.

Among other evidence \cite{Blake:2016wvh,Blake:2016jnn}, this is the pole-skipping phenomenon that is sometimes referred to as the smoking gun for the hydrodynamic origin of the chaotic mode \cite{Blake:2017ris}. Based on this, 
in the current paper and as the second part, we will explore the relation between hydrodynamics and quantum chaos in a many body system with chiral anomaly, via studying the pole-skipping phenomenon.

We mainly follow the argument of \cite{Blake:2018leo}\footnote{See also \cite{Grozdanov:2018kkt} for the first study of pole-skipping at finite coupling.} which is based on studying the near horizon dynamics of metric perturbations in Fourier space, namely $\delta g_{\mu \nu}(r_h;\omega, k)$'s.\footnote{Let us denote that by using symmetries, we only work with a subset of $\delta g_{\mu\nu}$'s.} To proceed,  the linearized form of one appropriate component of the Einstein equations, expanded near the horizon, is considered. 
Let us recall that in any system with Einstein gravity dual, we expect the Lyapunov exponent to be as $\lambda= 2 \pi T$. According to  \eqref{chaos_point}, it corresponds to the imaginary frequency $\omega^*=i \,2\pi T$. Precisely at this frequency, the above mentioned linearized component of the Einstein equations highly simplifies and leaves $\delta g_{vv}(r_h)$ decoupled from the other metric perturbations at the horizon\footnote{$v$ is the time coordinate in the Eddington-Finkelstein coordinates.}. At a general wave number $k$, $\delta g_{vv}(r_h)$ thus must vanish. For a set of sources of the stress tensor components on the boundary, the condition $\delta g_{vv}(r_h)=0$ actually picks up a unique regular solution for $\delta g_{\mu\nu}(r;\omega, k)$ in the bulk \cite{Blake:2018leo}\footnote{We    thank the anonymous referee for pointing out this to us.}.  

However, there exist two special wavenumbers $k^*$, at which, this equation is automatically satisfied. 
This is equivalent to saying that at the point $(\omega^*,k^*)$, fixing the boundary sources and imposing regularity at the horizon is
not sufficient to pick out a unique solution to $\delta g_{\mu \nu}(r;\omega, k)$ in the ingoing coordinates \cite{Blake:2018leo}. In order to find a solution in the bulk around this point, namely at a general point like $(r;\omega^*+\delta \omega,k^*+\delta k)$, in addition to the sources of the stress tensor, the value of the parameter $\delta\omega/\delta k$ is needed too.
Then by changing $\delta\omega/\delta k$, one can obtain a spectrum of solutions in the bulk, including both renormalizable and non-renormalizable solutions, and all passing through $(\omega^*,k^*)$. Let us take the corresponding values for renormalizable and non-normalizable solutions as $\left(\delta\omega/\delta k\right)_{(n)}$ and $\left(\delta\omega/\delta k\right)_{(nn)}$, respectively. In accordance with holographic dictionary,  
all points between $(\omega^*,k^*)$ and $(\omega^*+\delta \omega_{(n)},k^*+\delta k_{(n)})$, on the line with slope  $\left(\delta\omega/\delta k\right)_{(n)}$, are poles of $G^{\mathcal{R}}_{\epsilon\epsilon}$ in the boundary theory; similarly, all points between $(\omega^*,k^*)$ and $(\omega^*+\delta \omega_{(nn)},k^*+\delta k_{(nn)})$, on the line with slope $\left(\delta\omega/\delta k\right)_{(nn)}$,  are roots of $G^{\mathcal{R}}_{\epsilon\epsilon}$. Thus, both lines of poles and roots of  $G^{\mathcal{R}}_{\epsilon\epsilon}$ pass through $(\omega^*,k^*)$. As a result, the Green's function  becomes multi-valued at this point.
It is nothing but the pole-skipping, predicted by the effective field theory of \cite{Blake:2017ris}.  

Following the above discussion, we find the pole-skipping points of energy density Green's function in a system with chiral anomaly, via studying  the Einstein-Maxwell-Chern-Simons theory. It turns out that $\omega^{*}/k^{*}$ is exactly the same as butterfly velocity $v_{B}$ found from shock wave computations in the previous part. This coincidence might be an implicit evidence for hydrodynamic origin of quantum chaos in holographic anomalous systems.

Furthermore, it has
recently been shown that  the lack of information to uniquely define a correlation function is not specific to energy density correlation functions at chaos point; Green's functions of generic operators have also the same feature but at negative Matsubara frequencies and some appropriate
complex values of wavenumber \cite{Blake:2019otz}\footnote{See \cite{Natsuume:2019vcv,Wu:2019esr,Ahn:2019rnq, Li:2019bgc,Natsuume:2019sfp,Natsuume:2019xcy},  for some recent related works, specially \cite{Ceplak:2019ymw,Das:2019tga} for some exact results in AdS3 and 2-dim BCFT, respectively.}. The presence of such set of pole-skipping points in the lower half of the complex Fourier  plane shows that the dispersion relations of collective modes in boundary theory at  energy scales $\omega \sim T$ are directly constrained by the near horizon dynamics of bulk fields.

We use the same idea to compute the set of pole-skipping points  in the lower half plane by studying the dynamics of a scalar field near the horizon. We find the tower of pole-skipping points whose frequencies are actually given by the Matsubara frequencies $\omega_{\ell}=- i 2\pi T \ell; \,\,\ell=1, 2, \cdots$. It turns out that the spectrum is   deviated from the symmetric spectrum of a non-chiral system. 
In particular, we find that the sum of wavenumbers corresponding to pole-skipping points at $\ell^{th}$ Matsubara frequency, which we call it  $\Delta \boldsymbol{k}_{(\ell)}$, is a special quantity with a universal behavior. Its universality comes from the fact that it does not depend on the mass of scalar field and consequently neither does on the scaling dimension of the dual operator.  By computing this quantity at several Matsubara frequencies, we propose a closed formula for it, which gives the sum at a general Matsubara frequency. Interestingly, we show that $\Delta \boldsymbol{k}_{(\ell)}/\Delta \boldsymbol{k}_{c}=-\ell^2$, with $\Delta \boldsymbol{k}_{c}$ being the sum of wavenumbers corresponding to the chaos points. This suggests that  in a holographic anomalous  system, the quantity $\Delta \boldsymbol{k}_{(\ell)}$ can be regarded as another macroscopic manifestation of chiral anomaly.

In the rest of the paper, we firstly introduce the gravity set-up dual to the chiral system in 
\sec{set_up}. In \sec{sec_shock} by using the shock wave picture, we compute the butterfly velocities in longitudinal direction. In \sec{chiral_transport}, we derive the spectrum of hydro modes in the system under study and compare the results with the butterfly velocities. Then in \sec{sec_pole_skipping} we show that the pole-skipping in energy density correlators occurs at exactly the chaos points. In \sec{scalar_field} and its corresponding subsections, we study the pole-skipping points of boundary operator dual to a dynamical scalar field in the bulk. Finally, in \sec{conclusion} we end with reviewing our results and discussing about some follow-up directions.

\section{Holographic quantum chaos and pole-skipping in an anomalous system}
In this section we holographically investigate the relation between quantum chaos and chiral anomaly as well as that of quantum chaos and hydrodynamics, in an anomalous system. Following the detailed explanations in the Introduction, in what follows, we mostly focus on computations and corresponding results.
\subsection{Chiral system and its holographic dual: set up}
\label{set_up}
	The holographic dual of a chiral system in the presence of magnetic field is the Einstein-Maxwell-Chern-Simons theory. The action in the bulk is given by
	\begin{equation}\label{action}
	S =  \frac{1}{16 \pi G_5} \int_{\mathcal{M}} d^5 x \,\,\sqrt{-g} \left(R +\frac{12}{L^2} - F^{M N} F_{M N}\right)+S_{CS}+ S_{bdy}
	\end{equation}
	with the Chern-Simons action being as the following
	\begin{equation}\label{CS_action}
	S_{CS}=\frac{\kappa}{12 \pi G_5}\int A \wedge F \wedge F=\,\frac{\kappa}{48\pi G_5}\int d^5 x \sqrt{-g}\,\,\epsilon^{\rho \mu \nu \alpha \beta}A_{\rho}F_{\mu\nu}F_{\alpha \beta}.
	\end{equation}
	Here $S_{bdy}$ is the boundary counter term. The equations of motion  are given by:
	\begin{eqnarray}\label{Gauge_equ}
	\nabla_{\nu}F^{\nu\mu }+\frac{\kappa}{4} \epsilon^{\mu \nu \rho  \alpha\beta}F_{\nu \rho}F_{ \alpha \beta}&=&0\\\label{Einstein_equ}
	R_{\mu \nu}+4 g_{\mu \nu}+\frac{1}{3}F^{\alpha \beta}F_{\alpha \beta}\,\,g_{\mu \nu}+2 F_{\mu \rho} F^{\rho}_{\,\,\nu}&=&0
	\end{eqnarray}
	from which, one can find the following magnetized brane solution in the bulk
	\begin{equation}\label{metric}
	ds^2=\frac{dr^2}{f(r)}-f(r)dt^2+ e^{2W_T(r)}(dx_1^2+dx_2^2)+e^{2W_L(r)}(dx_3+C(r)dt)^2
	\end{equation}
	\begin{equation}\label{field_strenght}
	F=E(r) dr\wedge dt+B dx_1\wedge dx_2+ P(r) dx_3\wedge dr.
	\end{equation}
 The functions $f(r)$, $W_{L}(r)$, $W_{T}(r)$, $C(r)$, $E(r)$ and $P(r)$ were found analytically in \cite{DHoker:2009ixq}, in the limit $B\ll T^2$. Assuming  the following expansions for these functions 
\begin{align}
f&= f_0 + B^2 f_2 \,\,\,\,\,\,\,\,\,\,\,\,\,\,\,\,\,\,\,\,\,\,\,\,\,\,\,\,\,\,\,\,\,\,E = E_0 + B^2 E_2\nonumber\\\label{asymptotic}
W_L &= W_{L0} + B^2 W_{L2}\,\,\,\,\,\,\,\,\,\,\, \,\,\,\,\,\,\,\,\,\,\,\,\, C = C_0 + B C_1\\
W_T &= W_{T0} + B^2 W_{T2} \,\,\,\,\,\,\,\,\,\,\,\,\,\,\,\,\,\,\,\, \,\,\,P = P_0 + B P_1\nonumber
\end{align}
and by considering the zero order solutions
\begin{equation}\label{RN}
E_0= \frac{Q}{r^3},\,\,\,\,\,
W_{L0}=W_{T0}=\log(r),\,\,\,\,\, f_0= r^2 + \frac{Q^2}{3 r^4} - \frac{M}{r^2},
\end{equation}
the authors of \cite{DHoker:2009ixq} found the correction functions in \eqref{asymptotic} as some unevaluated integrals. Here $Q$ and $M$ can be written in terms of the horizons radii, $r_{\pm}$, which satisfy $f(r_{\pm})=0$:
\begin{align}
\frac{Q^2}{3}=r_+^2 r_-^2 (r_+^2+ r_-^2),\,\,\,\,M= r_+^4 +r_-^4 + r_+^2 r_-^2. 
\end{align}
The unevaluated integral functions found in \cite{DHoker:2009ixq} can be evaluated analytically in some regimes. In \cite{Abbasi:2018qzw}, all these functions have been computed in a double expansion over $B/T^2$ and $\mu/T$.  In the current paper we follow \cite{Abbasi:2018qzw} and work in the same limit. \footnote{In this limit, our results will be relevant to the magnetohydrodynamics of the boundary system in the presence of a constant background magnetic field  \cite{Grozdanov:2017kyl,Hernandez:2017mch}.} Here, however, we do not rewrite the explicit expressions of functions    \eqref{asymptotic} evaluated in \cite{Abbasi:2018qzw}. In order to study the chaos, we only need to know the near horizon limit of the  mentioned correction functions.
So in the following we just give their series expansion around the outer horizon, $r_{h}\equiv r_{+}$.
One formally writes
\begin{eqnarray}\nonumber
f(r)&=&f'(r_h)(r-r_h)+\frac{f''(r_h)}{2}(r-r_h)^2+\cdots\\
C(r)&=&C'(r_h)(r-r_h)+\frac{C''(r_h)}{2}(r-r_h)^2+\cdots
\end{eqnarray}
and 
\begin{eqnarray}\nonumber
W_{L,T}(r)&=&W_{L,T}(r_h)+W_{L}'(r_h)(r-r_h)+\cdots\\\nonumber
E(r)&=&E(r_h)+E'(r_h)(r-r_h)+\cdots\\\nonumber
P(r)&=&P(r_h)+P'(r_h)(r-r_h)+\cdots\nonumber
\end{eqnarray}
where the coefficients have been given in the Table 1 of the Appendix \ref{Tables}.

There is an important point about $r_h$ in the above expressions. In the absence of magnetic field, the location of horizon is the root of $f_0$ and is given by $r_{h}=\pi T$.
However, as discussed in \cite{Abbasi:2018qzw}, since $f_0$ gets a magnetic correction, the outer horizon and consequently both the Hawking temperature and the chemical potential of the solution have to be corrected as well. Considering the double expansion discussed above, the corrected outer horizon, is found to be as 
\begin{equation}
r_{h}=\pi  T\left[1+\frac{2}{3} \,\nu ^2 +\, \left(\frac{1}{6}+\frac{1}{9}\,\nu ^2  \left(3 \kappa ^2 (-3 \pi +6+\log (16))-4\right)\right)b^2\right]
\end{equation}
to second order in  
$\nu=\frac{\mu}{\pi T}$ and $b=\frac{B}{(\pi T)^2}$.

\subsection{Quantum chaos  and butterfly effect from holography}
\label{sec_shock}
Following the explanation about shock wave and its backreaction on the eternal black hole geometry in the Introduction, we first proceed to rewrite the metric \eqref{metric} and the field strength \eqref{field_strenght} in the Kruskal coordinates. To this end, by using the tortoise coordinate
\begin{equation}
dr^*=\frac{dr}{\sqrt{f(r)(f(r)-C(r)^2e^{2W_{L}(r)})}}
\end{equation}
and defining $\tilde{f}(r)=f(r)-C(r)^2e^{2W_{L}(r)}$, we find the Kruskal coordinates $U$ and $V$ in the left side of the Kruskal diagram as the following
\begin{equation}
\begin{split}
t+r^*=v&\,\,\,\,\,\,\,V=-e^{\frac{1}{2}\tilde{f}'(r_h)v}\\
t-r^*=u&\,\,\,\,\,\,\,U=e^{-\frac{1}{2}\tilde{f}'(r_h)u}.
\end{split}
\end{equation}
In terms of the Kruskal coordinates, metric may be rewritten as 
\begin{equation}\label{ds_past}
ds_{\text{past}}^2=A(U V) dU dV+B_{L}(U V)dx_3^2+B_{T}(U V)(dx_1^2+dx_2^2)+D(U V)\left(\frac{dV}{V}-\frac{dU}{U}\right) dx_3,
\end{equation}
where the functions $A$, $B$ and $D$ are given  by
\begin{equation}
A(U V)=\frac{4}{U V}\frac{ f(r)}{\tilde{f}'(r_h)},\,\,\,\,\,\,\,\,B_{L,T}(U V)=e^{2W_{L,T}(r)},\,\,\,\,\,\,\,\,\,D(U V)=\frac{2 C(r) }{\tilde{f}'(r_h)}e^{2W_{L}(r)}.
\end{equation}
Note that the subscript "past" is denoting that the metric corresponds to an eternal black hole in the past, before becoming perturbed by the shock wave. Similarly,  field strength must be rewritten in $U-V$ coordinates in the past. one writes
\begin{equation}\label{F_past}
F_{\text{past}}=G(U V) \,d V \wedge dU+ B \,dx_1 \wedge dx_2 + H(U V)\,dx_3 \wedge\left(\frac{dV}{V}+\frac{d U}{U}\right),
\end{equation}
with the functions $G$ and $H$ given by
\begin{equation}
G(U V)=-\frac{E(r)}{U V}\frac{\sqrt{f(r)(f(r)-C(r)^2e^{2W_{L}(r)})}}{\tilde{f}'(r_h)},\,\,\,\,\,\,\,\,H(U V)=P(r)\frac{\sqrt{f(r)(f(r)-C(r)^2e^{2W_{L}(r)})}}{\tilde{f}'(r_h)}.
\end{equation}
According to \cite{Shenker:2013pqb}, the backreaction of shock wave on the above background develops a "future" solution in the bulk which can be obtained by imposing the shift
$V\rightarrow V+ h(x)\delta (U)$ to the past one. Under this shift, \eqref{ds_past} and \eqref{F_past} take the following form:
\begin{eqnarray}\label{ds_fututre}
ds_{\text{future}}^2&=&ds_{\text{past}}^2-A(U V) h(x) \delta(U) \,dU^2 -\frac{D(U V)}{V} h(x)\delta(U)\,d Udx_3\\\label{F_fututre}
F_{\text{future}}&=&F_{\text{past}}-\,\frac{H(U V)}{V}h(x) \delta (U) \,dx_3 \wedge dU.
\end{eqnarray}
Plugging \eqref{ds_fututre} and \eqref{F_fututre} in \eqref{Einstein_equ} and considering the matter source in the right hand side of it as $T_{UU}\sim E e^{\frac{1}{2}\tilde{f}'(r_h)t_w}\,\delta (U)\delta^{3}(\vec{x})$ \footnote{$t_w$ is the past time at which the energy $E$ is released into the bulk.}, one finds that $h$ must obey the following equation near the horizon \footnote{Note that near the horizon $r=r_h(1- U V + O(U^2 V^2))$.}
\begin{equation}\label{Shock_equ}
\begin{split}
\bigg[\,A(0)\left(\frac{1}{B_{L}(0)}\partial_z^2+\frac{1}{B_{T}(0)}\partial_i^2-\frac{ D'(0)}{B_{T}(0)A(0)}\,\partial_z\right)\,\,\,\,\,\,\,\,\,\,\,\,\,\,\,\,\,\,\,\,\,\,\,\,\,\,\,\,\,\,\,\,\,\,\,\,\,\,\,\,\,\,\,\,\,\,\,\,\,\,\,\,\,\,\,\,\,\,\,\,\,\,\,\,\,\,\,\,\,\,\,\,\,\,\,\,\,\,\,\,\,\,\,\,\\
+\left(4\frac{A'(0)-2A(0)^2}{A(0)}+2(2W_T'(0)+W'_{L}(0))\right)\,\,\,\,\,\,\,\,\,\,\,\,\,\,\,\,\,\,\,\,\,\,\,\,\,\,\,\,\,\,\,\,\,\,\,\,\,\,\,\,\,\,\,\,\,\,\,\,\,\,\,\,\,\,\,\,\,\,\,\,\,\,\,\,\,\,\,\,\,\,\\
-\frac{128\,r_h^2}{3\tilde{f}'(r_h)^2}\frac{E(0)^2}{A(0)}-\frac{4}{3}\frac{A(0)}{B_{T}(0)^2}B^2+\,\frac{D'(0)^2}{B_{T}(0)A(0)}\bigg]h(x)\sim 2  E e^{\frac{1}{2}\tilde{f}'(r_h)t_w}\,\delta^3(\vec{x})
\end{split}
\end{equation}
where $E(0)=E(UV)|_{U=0}\equiv E(r)|_{r=r_h}$. The same convention has been considered for functions $W_T$ and $W_L$ in the above equation. 

Note that through deriving the equation above we have already taken the magnetic field in the third direction. For a magnetic field in a general direction, say $\vec{b}$, equation \eqref{Shock_equ} can be formally written as the following 
\begin{equation}\label{formal_shock_equation}
\left( \partial_{\parallel}^2+{\textswab{q}}^2\frac{}{} \partial _{\perp}^2 +2 \,{\textswab{p}}\, \hat{b}\cdot \vec{\partial}- m_0^2\right)h(x)\sim \frac{2 B_{L}(0)}{A(0)} E e^{\frac{1}{2}\tilde{f}'(r_h)t_w}\,\delta^3(\vec{x})
\end{equation}
where $\parallel$ and $\perp$ are denoting the directions parallel and perpendicular to the magnetic field, respectively. It should be noted that $b$ in the third term of the left hand side is coming from $D'(0)$ in the first line of  \eqref{Shock_equ}.  It is also obvious that ${\textswab{q}}$, ${\textswab{p}}$ and $m_0$ can be easily read in terms of coefficients in \eqref{Shock_equ}.

When $h$ is assumed to be only function of $x_3$, the second term in \eqref{formal_shock_equation} vanishes and $h(x_3)$ is found to be (See Appendix \ref{App_contour} for detail of computations.)\footnote{$t_{*}$ is the scrambling time \cite{Roberts:2014isa}.}
\begin{equation}\label{butterfly_equ_x_3}
\begin{split}
h(x_3)\sim& \,-\pi E\, e^{\frac{1}{2}\tilde{f}'(r_h)(t_\omega-t_*)-\left({\textswab{p}}+\sqrt{{\textswab{p}}^2+m_0^2}\right)x_3 }\,\,\sqrt{{\textswab{p}}^2+m_0^2}\,\,\theta(x_3)\\
&\,\,\,\,-\pi E\, e^{\frac{1}{2}\tilde{f}'(r_h)(t_\omega-t_*)-\left(-{\textswab{p}}+\sqrt{{\textswab{p}}^2+m_0^2}\right)x_3 }\,\,\sqrt{{\textswab{p}}^2+m_0^2}\,\,\theta(-x_3).
\end{split}
\end{equation}
Comparing this equation with \eqref{OTOC} and \eqref{chaos_point}, one finds the Lyapunov exponent of the system, as expected, is 
\begin{equation}
\lambda=\frac{1}{2}\tilde{f}'(r_h)=\, 2 \pi T.
\end{equation}
Moreover, we see that the speed at which the information propagates depends on the sign of $x_3$; let us recall that in \eqref{OTOC}, at $t=0$, the operator $W$ is inserted at $x=0$ and then at time $t$,   $V$ will be inserted at $\vec{x}$. Thus $x_3>0$ corresponds to the propagation of information in $+x_3$ direction, namely the magnetic field direction;
similarly $x_3<0$ corresponds to the butterfly propagation opposite to the magnetic field.  From \eqref{butterfly_equ_x_3}, we find the corresponding butterfly velocities as the following
\begin{equation}\label{butterfly_from_shcok}
\begin{split}
x_3>0:&\,\,\,\,\,\,\,\,\,\,\,\,\,v_{B}^{L_1}=\frac{2\pi T}{m_0^2}\left(\sqrt{{\textswab{p}}^2+m_0^2}-{\textswab{p}}\right)\\
x_3<0:&\,\,\,\,\,\,\,\,\,\,\,\,\,v_{B}^{L_2}=-\frac{2\pi T}{m_0^2}\left(\sqrt{{\textswab{p}}^2+m_0^2}+{\textswab{p}}\right).
\end{split}
\end{equation}
When the magnetic field vanishes, ${\textswab{p}}=0$ and one finds the degenerate velocities of \cite{Blake:2016wvh}.  Interestingly, the above relations show that when ${\textswab{p}}\ne0$, the speed of butterfly velocity in the medium is not isotropic. In the following we show that this asymmetry is just coming from the chiral anomaly.

To better explain the physical aspects of this result, we first compute  ${\textswab{p}}$ and $m_0$ in terms of background data. We find:
\begin{equation}\label{p_m_o}
\begin{split}
{\textswab{p}}&= (\pi T)\, \kappa \,(\log(4)-1) \nu^2 \,b\\
m_0^2&= (\pi T)^2 \left[6+36 \nu^2-\left(\frac{\pi^2}{6}-1\right)b^2-\left(\pi^2+\frac{92}{9}+56 \kappa^2 (\log (2)-1)\right)\nu^2b^2\right]
\end{split}
\end{equation}
Now we can put these expressions back into \eqref{butterfly_from_shcok} and read the butterfly velocities. In the following we discuss on the resultant formulas in various limits.
\newline
$\bullet$ \textbf{Uncharged system in the absence of magnetic field}

Such system is holographically described with a Schwarzschild black brane. Substituting $b=0$ and $\nu=0$ in \eqref{butterfly_from_shcok}, we reproduce the well known result $v_{B}=\sqrt{2/3}$ \cite{Shenker:2013pqb} for this system.
\newline
$\bullet$ \textbf{Charged system in the absence of magnetic field}

The holographic dual of this system is a Reissner-Nordstorm black brane. Substituting $b=0$,  in \eqref{butterfly_from_shcok}, we reproduce the result of \cite{Blake:2016jnn} concerning the RN black brane as the following
\begin{equation}
v_{B}^{L_{1,2}}=\pm v_{B}\left(1-\frac{\mu^2}{3\pi^2 T^2}\right)
\end{equation}
We study chaos in RN black brane at finite $\mu$ in a future work, in details \cite{Abbasi}.
\newline
$\bullet$ \textbf{Non-chirally charged system in the presence of magnetic field}

Such system is holographically dual to a magnetized RN black brane. In order to put off the chiral effects in \eqref{butterfly_from_shcok}, we force the Chern-Simons coefficient to vanish, i.e. $\kappa =0$. As a result, ${\textswab{p}}$  in \eqref{p_m_o} vanishes and then we obtain two equal butterfly velocities with opposite signs:
\begin{equation}\label{v_long_non_chiral}
v_{B, \kappa=0}^{L_{1,2}}=\pm v_{B}\left[1-\frac{\mu^2}{3(\pi T)^2}+\left(\frac{\pi^2-6}{72}\right)\,\,\frac{B^2}{(\pi T)^4}-\left(\frac{\pi^2+54}{216}\right)\,\,\frac{\mu^2 B^2}{(\pi T)^6}\right]
\end{equation}
In \cite{Blake:2017qgd}, based on shock wave picture of \cite{Shenker:2013pqb}, the anisotropic butterfly velocities for a general family of anisotropic metrics has been computed. Our metric given in \eqref{metric} for $C(r)=0$, namely for the non-chiral case, falls into the family of metrics studied in \cite{Blake:2017qgd}. In the Appendix \ref{Blake_Davison_Sachdev}, we will show that our above result in \eqref{v_long_non_chiral} is in complete agreement with that of \cite{Blake:2017qgd} and \cite{Ling:2016ibq}.
\newline
$\bullet$ \textbf{Chirally charged system in the presence of magnetic field}

Our main goal in this section is to just consider this case. Since $\kappa$ is proportional to the chiral anomaly coefficient in a chiral system, we take $\kappa\ne0$ and find
\begin{equation}\label{v_B_L}
\begin{split}
v_{B}^{L_{1,2}}=\pm\sqrt{\frac{2}{3}}\left(1-\frac{\mu^2}{3(\pi T)^2}\right)&-\frac{2}{3}\,\kappa\, (\log (4)-1) \frac{\mu^2 B}{(\pi T)^4}\\
\pm &\left(\frac{\pi ^2-6}{36 \sqrt{6}}-\frac{ \pi^2+18(-4 \kappa^2(\log(4)-2))}{108 \sqrt{6}}\frac{\mu^2}{(\pi T)^2}\right)\frac{B^2}{(\pi T)^4} 
\end{split}
\end{equation}
It is obvious that $\mu^2 B$ term splits the degeneracy between the magnitude of two butterfly velocities in this case, so the equality $v_{B}^{L_1}=|v_{B}^{L_2}|$ will no longer hold. Putting the value of $\kappa$ \cite{Cvetic:1999ne,Chamblin:1999tk}, we find
\begin{equation}\label{deltav}
\kappa=-\frac{2}{\sqrt{3}}\,\,\,\,\,\rightarrow\,\,\,\,\,\boxed{
	\Delta v_{B}^{L}=v_{B}^{L_1}-|v_{B}^{L_2}|=\,\frac{8}{3\sqrt{3}}(\log 4-1)\, \frac{\mu^2\,B}{(\pi\,T)^4}}
\end{equation}
This is actually our central result in this section. It is seen that the difference between the magnitude of the right- and left-moving butterfly propagation is originated from the anomaly of microscopic quantum field theory. Consequently, any macroscopic observation of such difference in experiment is a new manifestation of chiral anomaly.

For later requirements, we know represent the butterfly velocities \eqref{v_B_L} in the form of chaos points introduced in \eqref{chaos_point}. The chaos points are given by $(i \omega_{c}, i k_{c})$ where $\omega_{c}=2\pi T$ and $k_{c}= 2\pi T \boldsymbol{k}_c$. $\boldsymbol{k}_c$  is given by
\begin{equation}\label{k_long_chaos_1_2}
\boldsymbol{k}_{c;1,2}=\pm\frac{\sqrt{6}}{2}\pm\frac{\sqrt{6}}{6}\nu ^2+ \,\kappa\,(\log (4)-1)\,b   \nu ^2\pm\left(\frac{6-\pi ^2}{24 \sqrt{6}}+\frac{ 66-\pi^2-72 \kappa^2(\log(4)-2)}{72 \sqrt{6}}\nu ^2\right)b^2,
\end{equation}
We also call the sum of the above two wavenumbers as $\Delta \boldsymbol{k}_{c}$
\begin{equation}\label{k_chaos}
\Delta \boldsymbol{k}_{c}\equiv\boldsymbol{k}_{c;1}+\boldsymbol{k}_{c;2}=\,2\kappa\,(\log (4)-1)\, \nu ^2b.
\end{equation}
Before ending this section, let us note that 
we will compute the butterfly velocities in the transverse directions in the Appendix \ref{App_transverse}.

\subsection{Butterfly effect and chiral transport}
\label{chiral_transport}
Hydrodynamics is the universal low-energy long wavelength limit of thermal systems. Instead of microscopic degrees of freedom, hydrodynamics uses  just a few number of macroscopic variables as the dynamical degrees of freedom, say velocity $u^{\mu}(t,\vec{x})$, temperature $T(t,\vec{x})$ and $\cdots$. The idea of hydrodynamics is that in the long wavelength limit, every physical quantity in the system, like energy density, can be written in a derivative expansion of the above degrees of freedom. The equations of motion are then simply the (non-)conservation equations of energy-momentum tensor (and global currents). In a simple system with one anomalous $U(1)$ current, one writes
\begin{eqnarray}\label{hydro_eq1}
\partial_{\mu}T^{\mu \nu}&=&F^{\nu}_{\,\,\,\rho}J^{\,\rho}\\
\partial_{\mu} J^{\mu}&=&C E^{\mu}B_{\mu}
\end{eqnarray}
with the electric and magnetic fields in the rest frame of fluid elements defined as $E^{\mu}=F^{\mu \nu}u_{\nu}$ and  $B^{\mu}=\frac{1}{2}\epsilon^{\mu\nu\alpha\beta}u_{\nu}F_{\alpha \beta}$, respectively. $C$ is the coefficient of chiral anomaly in the system.
In a chiral system and in presence of the background field $F_{\mu\nu}$, the constitutive relations are given by \cite{Son:2009tf,Landsteiner:2012kd}
\begin{eqnarray}
\label{Tmunu}\label{T}
T^{\mu \nu} &= & w u^{\mu} u^{\nu} + p g^{\mu \nu}+\sigma^{\mathcal{B}}_{\epsilon}(u^{\mu}B^{\nu}+u^{\nu}B^{\mu})+\sigma^{\mathcal{V}}_{\epsilon}(u^{\mu}\omega^{\nu}+u^{\nu}\omega^{\mu}),\\\label{J}
\label{jmu}
J^{\mu} & = & nu^{\mu} +\sigma^{\mathcal{B}} B^{\mu}+ \sigma^{\mathcal{V}} \omega^{\mu},
\end{eqnarray}
where $\omega^{\mu}=\epsilon^{\mu\nu\alpha\beta}u_{\nu}\partial_{\alpha}u_{\beta}$ and $\sigma^{\mathcal{B}}$, $\sigma^{\mathcal{B}}_{\epsilon}$, $\sigma^{\mathcal{V}}$ and $\sigma^{\mathcal{V}}_{\epsilon}$ are the anomalous transport coefficients. Here also, $w=\epsilon+p$ is the enthalpy density of the equilibrium.

Hydrodynamic excitations are gap-less modes of the equation \eqref{hydro_eq1} and \eqref{hydro_eq1} propagating on top of the thermal equilibrium state. In order to find them, we take the following near equilibrium profile for the hydrodynamic variables
\begin{eqnarray}\nonumber
u^{\mu}(t,x_3)&=&(1,0,0,0)+\bigg(0,\delta u_1(t,x_3),\delta u_2(t,x_3),\delta u_{3}(t,x_3)\bigg)\\
T(t,x_3)&=&T+\delta T(t,x_3)\\\nonumber
\mu(t,x_3)&=&\mu+\delta \mu(t,x_3).
\end{eqnarray}
We have  assumed the time dependent perturbations are  functions of only the third spacial coordinate. We also take the magnetic field being directed along the same axis. Based on these assumptions, we will find the plane wave hydrodynamic excitations propagating either parallel or opposite to the magnetic field.  Let us recall that around the equilibrium state, the thermodynamic quantities deviate from their equilibrium values, too. One writers
\begin{eqnarray}\label{delta_epsilon}\nonumber
\delta \epsilon(t,x_3)&=&\epsilon+\left(\frac{\partial \epsilon}{\partial T}\right)_{\mu}\, \delta T(t,x_3)+\left(\frac{\partial \epsilon}{\partial \mu}\right)_{T}\, \delta \mu(t,x_3)\,\equiv \,\epsilon+\alpha_1 \delta T+\,\alpha_2 \delta \mu\\\label{delta_pressure}
\delta p(t,x_3)&=&p+\left(\frac{\partial p}{\partial T}\right)_{\mu}\, \delta T(t,x_3)+\left(\frac{\partial p}{\partial \mu}\right)_{T}\, \delta \mu(t,x_3)\,\equiv\, p+\gamma_1 \delta T+\,\gamma_2 \delta \mu\\\label{delta_charge}\nonumber
\delta n(t,x_3)&=&n+\left(\frac{\partial n}{\partial T}\right)_{\mu}\, \delta T(t,x_3)+\left(\frac{\partial n}{\partial \mu}\right)_{T}\, \delta \mu(t,x_3)\,\equiv \,n+\beta_1 \delta T+\,\beta_2 \delta \mu.
\end{eqnarray}
Expanding the equations of motion to first order in the perturbations and also to second order in derivatives, then, results in the following five linear equations:
\begin{eqnarray}\nonumber
0&=&\big(\alpha_1\, \omega-\partial_{T}\sigma^{\mathcal{B}}_{\epsilon} \,B k\big)\delta T+\big(\alpha_2\, \omega-\partial_{\mu}\sigma^{\mathcal{B}}_{\epsilon} \,B k\big)\frac{\delta \mu}{w}+\big(- w k-2\sigma^{\mathcal{B}}_{\epsilon} \,B \omega\big)\delta u_3\\\nonumber
0&=&i\big(n B+\sigma_{\epsilon}^{\mathcal{V}}k \omega\big)\delta u_2- w \omega \,\delta u_1\\
0&=&-i\big(n B+\sigma_{\epsilon}^{\mathcal{V}}k \omega\big)\delta u_1- w \omega \,\delta u_2\\\nonumber
0&=&\big(\gamma_1\, k-\partial_{T}\sigma^{\mathcal{B}} \,B \omega\big)\delta T+\big(\gamma_2\, k-\partial_{\mu}\sigma^{\mathcal{B}_{\epsilon}} \,B \omega\big)\frac{\delta \mu}{w}+\big(- w \omega+2\sigma^{\mathcal{B}}_{\epsilon} \,B k\big)\delta u_3\\\nonumber
0&=&\big(\beta_1\, \omega-\partial_{T}\sigma^{\mathcal{B}} \,B k\big)\delta T+\big(\beta_2\, \omega-\partial_{\mu}\sigma^{\mathcal{B}} \,B k\big)\frac{\delta \mu}{w}+\big(- n k+\sigma^{\mathcal{B}} \,B \omega\big)\delta u_3
\end{eqnarray}
The eigen modes of the above equations are the so-called hydrodynamic modes. In this case with only one single $U(1)$ anomalous currents, one finds five of them, including two sound waves, two chiral Alfv\'en waves (CAW) together with one chiral magnetic-heat wave (CMHW) \cite{Abbasi:2015saa,Ammon:2017ded} \footnote{See \cite{Rybalka:2018uzh} for a related work.}. In the laboratory frame \cite{Abbasi:2017tea}, we find the velocity of these modes as the following:
\begin{eqnarray}\label{v_CAW}
v_{CAW}&=&\frac{n\,B}{w^2}\,\sigma^{\mathcal{V}}_{\epsilon}\\\label{v_CMHW}
v_{CMHW}&=&\frac{B}{w}\frac{1}{[\beta, \alpha]}\bigg(w\big(\alpha_1 \partial_{\mu}+\alpha_2 \partial_{T}\big)\sigma^{\mathcal{B}}-n\big(\alpha_1 \partial_{\mu}+\alpha_2 \partial_{T}\big)\sigma^{\mathcal{B}}_{\epsilon}\bigg)\\\label{v_sound}
v_{sound}&=&\pm c_s+\frac{B}{2w}\frac{[\gamma,\alpha]}{[\beta, \alpha]}\left(-1+\frac{(n\alpha_2-w\beta_2)\partial_{T}-(n\alpha_1-w\beta_1)\partial_{\mu}}{n[\gamma,\alpha]-w[\gamma, \beta]}\right)\sigma^{\mathcal{B}}\\\nonumber
&&+\frac{B}{w}\left(1-\frac{[\gamma, \beta]}{[\alpha, \beta]}\right)\sigma^{\mathcal{B}}_{\epsilon}+\,\frac{B}{2w}\frac{[\gamma,\alpha]}{[\beta, \alpha]}\left(\frac{(n\alpha_1-w\beta_1)\partial_{\mu}-(n\alpha_2-w\beta_2)\partial_{T}}{n[\gamma,\alpha]-w[\gamma, \beta]}\right)\sigma^{\mathcal{B}}_{\epsilon}
\end{eqnarray}
where $c_s^2=\frac{[\beta,\gamma]}{[\beta, \alpha]}+\frac{n}{w}\frac{[\gamma,\alpha]}{[\beta, \alpha]}$. We have used the shorthand notation $[\alpha,\beta]=\alpha_1 \beta_2-\alpha_2 \beta_1$, and  similarly we use the same for other commutators.

Our results regarding the hydrodynamic modes have been general so far, in the sense that  we used no any special equation of state to find them. In order to evaluate the above velocities in a specific system with a given equation of state, we have to firstly compute the corresponding thermodynamic derivatives $\alpha_{1,2}$, $\beta_{1,2}$ and $\gamma_{1,2}$ of the system. 
Let us recall that the system under study in the current paper is a holographic chiral system dual to Einstein-Maxwell-Chern-Simons theory, \eqref{action} and \eqref{CS_action}. 
The  thermodynamic equation of state for such system has been found in \cite{Abbasi:2018qzw}. The energy density, thermodynamic pressure and charge density are to second order in $\nu$ and $b$ given by
\begin{equation}\label{T00_a_3_final}
\begin{split}
&\epsilon=\frac{N_c^2}{8\pi^2}\big(3(\pi T)^4+12 (\pi T)^2\mu^2+8 \mu^4\big)+\frac{N_c^2B^2}{4 \pi^2}\bigg((1- \log( \frac{\pi T}{\Delta}))-\frac{2}{3}\frac{\mu^2}{\pi T^2}(8 \log (2)-3)\bigg)\\
&p=\frac{N_c^2}{24\pi^2}\big(3(\pi T)^4+12 (\pi T)^2\mu^2+8 \mu^4\big)+\frac{N_c^2B^2}{4 \pi^2}\bigg( \log( \frac{\pi T}{\Delta})+\frac{2}{3}\frac{\mu^2}{\pi T^2}(8 \log (2)-3)\bigg)\\
&n = \frac{N_c^2}{3\pi^2}\big(3(\pi T)^2\,\mu+4 \mu^3\big)+\frac{N_c^2B^2}{3\pi^2}\frac{\mu}{(\pi T)^2}\left(8 \log(2)-3\right)
\end{split}
\end{equation}
where $N_c\gg1$ is the number of colors in the boundary gauge theory and $\Delta$ is an energy scale.
Using these expressions, one then can simply compute the thermodynamic derivatives defined in \eqref{delta_epsilon}, \eqref{delta_pressure} and \eqref{delta_charge} (See Appendix \ref{thermo_derivatives}.). 

The only things which then remain to be specified are the anomalous transport coefficients.\footnote{For a general system with $U(1)$ triangle anomalies some expressions have been found for these coefficients in \cite{Son:2009tf}. Analogue of the expressions found in \cite{Son:2009tf} for a system with anomalous $SU(N)$ symmetries,  was found in \cite{Neiman:2010zi}. Both the mentioned references computed the anomalous transport coefficients in the Landau-Lifshitz frame. For the expressions in the laboratory frame, see \cite{Landsteiner:2012kd,Abbasi:2017tea}.}
The non-dissipative transport coefficients introduced in \eqref{T} and \eqref{J} have been computed for the Einstein-Maxwell-Chern-Simons theory in \cite{Landsteiner:2012kd}. However in that reference, in addition to the Chern-Simons coupling \eqref{CS_action}, a gauge-gravitational coupling has been considered as well:
\begin{equation}
S_{CS}=\frac{1}{16\pi G}\int d^5x \sqrt{-g}\,\,\epsilon^{\rho \mu \nu \alpha \beta}A_{\rho}\left(\frac{\kappa}{3}F_{\mu \nu}F_{\alpha \beta}+\frac{\lambda}{4}R^{\sigma}_{\,\,\eta\mu\nu}R^{\eta}_{\,\,\sigma\alpha \beta}\right)
\end{equation}
where $\lambda$ is proportional to the gravitational anomaly coefficient in the boundary theory and as mentioned earlier, $\kappa$ is proportional to the coefficient of chiral anomaly. It is clear that in order to use the results of \cite{Landsteiner:2012kd} in our work, we have to take the limit $\lambda=0$. Doing so, the anomalous transport  coefficients take the following simple form 
\begin{eqnarray}\nonumber
\sigma^{\mathcal{B}}&=&-\frac{\kappa}{2 \pi G_5}\,\mu=\,\frac{2N_c^2}{\pi^2\sqrt{3}}\,\mu\\
\sigma^{\mathcal{V}}=\sigma^{\mathcal{B}}_{\epsilon}&=&-\frac{\kappa}{2 \pi G_5}\,\mu^2=\,\frac{N_c^2}{\pi^2\sqrt{3}}\,\mu^2\\\nonumber
\sigma^{\mathcal{V}}_{\epsilon}&=&-\frac{\kappa}{6 \pi G_5}\,\mu^3=\,\frac{2N_c^2}{3\pi^2\sqrt{3}}\,\mu^3
\end{eqnarray}
where in the second equalities we have used \cite{Cvetic:1999ne,Chamblin:1999tk} 
\begin{equation}\label{C_coef}
C=-\frac{\kappa}{2\pi G_5}=\,\frac{2 N_c^2}{\pi^2 \sqrt
	3}.
\end{equation}
Putting all the above expressions into the velocities \eqref{v_CAW}, \eqref{v_CMHW} and \eqref{v_sound} and expanding to second order in $b$ and $\nu$, we find
\begin{equation}\label{velocities}
\begin{split}
v_{sound}=&\,\frac{1}{\sqrt{3}}\left(\pm1+\frac{4}{3 }\nu^2 b\right)\\
v_{\text{CMW}}=&\,\frac{2}{\sqrt{3}}b\left(1-2\nu^2\right)\\
v_{\text{CAW}}=&\,0.
\end{split}
\end{equation}
There are some points regarding the above results which deserve more explanation. Firstly, the non-propagation of chiral Alfv\'en wave in this case shows that these kind of gap-less modes is tied to the presence of gravitational anomaly. If we took $\lambda\ne0$, we would obtain $v_{CAW}\ne0$, although just in the laboratory frame (\cite{Yamamoto:2015ria,Abbasi:2017tea}).
The second point is that in front of every magnetic field factor $b$, there is an implicit chiral anomaly coefficient $C$. But since our computations are in the framework of holography, the relation \eqref{C_coef} has been already imposed and the anomaly coefficient is not seen explicitly.

More importantly, from \eqref{velocities} one simply notices that the right- and left-moving sound modes have different velocities $v_{sound,1}$ and $v_{sound,2}$.
The difference between their magnitudes is
\begin{equation}
\Delta v_{sound}=\,\frac{8}{3\sqrt{3}}\nu^2 b.
\end{equation}
 It can be simply seen that $\Delta v_{sound}$ arises just due to the non-zero chiral magnetic effect coefficients $\sigma^{\mathcal{B}}$ and $\sigma^{\mathcal{B}}_{\epsilon}$, in \eqref{v_sound}.
Comparing this result with $\Delta v_{B}$ (see \eqref{deltav}), one finds
\begin{equation}\label{chaos_hydro}
	\frac{\Delta v_{B}^{L}}{\Delta v_{\text{sound}}}=\,\frac{(\log 4 -1)8/3\sqrt{3}}{8/3\sqrt{3}}=\,\log 4-1.
\end{equation}
This equation is not an exact relation between $\Delta v_{B}^{L}$ and $\Delta v_{sound}$. We have just shown that it holds perturbatively, up to second order in $\nu$ and $b$. 
If one shows that this equation continues to hold beyond the perturbative limit of small $\nu$ and $b$, then it will send a physical message: the splitting between the butterfly velocities not only  will be the manifestation of chiral anomaly, but also will reveal the presence of chiral magnetic effect. Then in addition to previously proposed macroscopic probes of the the chiral magnetic effect, like three particle correlations in heavy-ion collisions \cite{Abelev:2009ac},  $\Delta v_{B}^{L}$ might be regarded as a new experimental probe of this effect in anomalous systems. We leave the study on the non-perturbative regime of $\nu$ and $b$ to a future work.

\subsection{Pole-skipping of energy density Green's function from holography }
\label{sec_pole_skipping}
In order to find the pole-skipping points of energy density Green's function, we have to work in the ingoing Eddington-Finkelstein coordinates . In these coordinates, with time coordinate identified by $v$,\footnote{The time coordinate $v$ is defined as 
	\begin{equation*}v=t+\int\frac{dr}{\sqrt{f(r)\left(f(r)-C(r)^2e^{2W_{L}(r)}\right)}}	\end{equation*} in terms of coordinates $r$ and $t$ in \eqref{metric}. In the Eddington-Finkelstein coordinates, metric and field strength are non-singular everywhere away from $r=0$.} equations \eqref{metric} and  \eqref{field_strenght} take the following form
\begin{eqnarray}\label{Eddington}
\begin{split}
ds^2=&-F(r)dv^2+2q(r)dr dv+2\left(j(r) dv +\frac{}{}s(r)dr\right) dx_3+e^{2W_{T}(r)}(dx_1^2+dx_2^2)+e^{2W_{L}(r)}dx_3^2\\
F=&\,E(r) dr\wedge dv+B dx_1\wedge dx_2+ P(r) dx_3\wedge dr 
\end{split}
\end{eqnarray}
where
\begin{eqnarray}\nonumber
F(r)&=&f(r)-e^{2W_{L}(r)}C(r)^2\\
q(r)&=&\left(1-e^{2W_{L}(r)}\frac{C(r)^2}{f(r)}\right)^{1/2}\\\nonumber
j(r)&=&C(r)e^{2W_{L}(r)}\\\nonumber
s(r)&=&-\frac{C(r)e^{2W_{L}(r)}}{\sqrt{f(r)\left(f(r)-C(r)^2e^{2W_{L}(r)}\right)}}.
\end{eqnarray}

In order to find the energy density two-point function, one has to consider the perturbations of the $vv$-component of metric $\delta g_{vv}(r,v,x)=\delta g_{vv}(r)e^{-i\omega v+i \vec{k}\cdot \vec{x}}$. It is clear that among the other Einstein equations, the $vv$ component is the one which governs the dynamics of $\delta g_{vv}$. Thus we study the linearized form of this equation in two cases;  while we always consider the magnetic field being directed along the third direction, we take once, the perturbations propagating in the same direction and then as the second case, we study the propagation of the transverse direction (See Appendix \ref{App_transverse} for the transverse case.). 
For either case, we find the points $(\omega^*, k^*)$ at which the $vv$ equation becomes trivial. It is in correspondence with the multi-valuedness  point of the boundary energy two-point function. 

In the longitudinal case, perturbations $\delta g_{vv}$, $\delta g_{rr}$, $\delta g_{rv}$, $\delta g_{x^ix^i}$, $\delta g_{x^3x^3}$, $\delta g_{vx^3}$ and $\delta g_{rx^3}$ decouple form the others. Thus just by considering these metric perturbations around \eqref{Eddington}, the linearized $vv$ component of Einstein equations near the horizon reads  (up to second order in $b$ and $\nu$):
\begin{equation}\label{NH_linear_Long}
\begin{split}
&\bigg[	k^2-3i \pi T \omega+\left(-\frac{4}{3}k^2+2i \pi T \omega\right)\nu^2+\left(\big(\frac{\pi^2}{36}-\frac{1}{3}\big) k^2+\frac{1}{2}i \pi T \omega\right)b^2+\mathcal{A}\,\nu^2b+\mathcal{B}\,\nu^2b^2\bigg]\delta g_{vv}^{(0)}\\
&\,\,\,\,\,\,\,\,\,\,\,\,\,\,\,\,\,\,\,\,\,\,\,\,\,\,\,\,\,\,\,\,\,\,\,\,\,\,\,\,	-i (2\pi T+i \omega)\,\frac{4\nu^2-3}{216}\,\bigg[ -2 (36+(\pi^2-12)b^2)\bigg(2k\,\delta g_{vx^3}^{(0)}+\omega \,\delta g_{x^3x^3}^{(0)}\bigg)\\
&\,\,\,\,\,\,\,\,\,\,\,\,\,\,\,\,\,\,\,\,\,\,\,\,\,\,\,\,\,\,\,\,\,\,\,\,\,\,\,\,\,\,\,\,\,\,\,\,\,\,\,\,\,\,\,\,\,\,\,\,\,\,\,\,\,\,\,\,\,\,\,\,\,\,\,\,\,\,\,\,\,\,\,\,\,\,\,\,\,\,\,\,\,\,\,\,\,\,\,\,\,\,\,\,\,\,\,\,\,\,\,\,\,\,\,\,\,\,\,\,\,\,\,\,\,\,\,\,\,\,\,\,\,\,\,\,+\,\omega\,\left(-72+(\pi^2+24)b^2\right)\,\delta g_{x^ix^i}^{(0)}\bigg]=\,0
	\end{split}
\end{equation}
Where the superscript $(0)$  refers to the zeroth order in the expansion around $r=r_h$. The coefficients $\mathcal{A}$ and $\mathcal{B}$ are given by
\begin{eqnarray}
\mathcal{A}&=&-4 i \pi T\, \kappa \,(\log(4)-1)\, k\, \nu^2 b\\\nonumber
\mathcal{B}&=&2 i \pi T \,\bigg[(\log(4)-2)\kappa^2-1\bigg]\, \omega\, \nu^2 b^2+\frac{12-\pi^2}{27}k^2\, \nu^2 b^2\
\end{eqnarray}
Let us denote that as one expects, when $\nu=0$ and $b=0$, equation \eqref{NH_linear_Long} simplifies to what was found for the Schwarzschild case in \cite{Blake:2018leo}. Obviously, when $\omega =i \omega^{*}=i 2 \pi T $, all the other fields decouple from $\delta g_{vv}$  at the horizon and we are left with
\begin{equation}\label{delta_g_vv}
	\bigg[	k^2+\,6 \pi^2 T^2+\left(-\frac{4}{3}k^2-4 \pi^2 T^2 \right)\nu^2+\left(k^2\big(\frac{\pi^2}{36}-\frac{1}{3}\big)-\, \pi^2 T^2 \right)b^2+\mathcal{C}\,\nu^2b+\mathcal{D}^{*}\,\nu^2b^2\bigg]\delta g_{vv}^{(0)}=0
\end{equation}
where $\mathcal{D}^{*}=\mathcal{D}(i\, 2\pi T, k)$. For a generic $k$, the above equation gives $\delta g_{vv}^{(0)}=0$, however, for some special value of $k=i k^{*}$, this equation is automatically satisfied. Such $k^{*}$ can be computed perturbatively in a double expansion over $b$ and $\nu$;  considering $k^{*}= 2 \pi T \boldsymbol{k}^*$,  one finds the following two dimensionless wavenumbers:
\begin{equation}\label{k_pole_skipping_1_2}
\boldsymbol{k}^{*}_{1,2}=\pm\frac{\sqrt{6}}{2}\pm\frac{\sqrt{6}}{6}\nu ^2+ \,\kappa\,(\log (4)-1)\,b   \nu ^2\pm\left(\frac{6-\pi ^2}{24 \sqrt{6}}+\frac{ 66-\pi^2-72 \kappa^2(\log(4)-2)}{72 \sqrt{6}}\nu ^2\right)b^2,
\end{equation}
As it is obvious, this result is exactly the same as \eqref{k_long_chaos_1_2}. In other words, we have shown that in an anomalous system, the pole-skipping points of energy density Green's function coincides with the chaos points: 
\begin{equation}\label{coinsidence}
(\omega_c, k_c)\equiv\,(\omega^{*}, k^{*})
\end{equation}
This results might be regraded as the first evidence for the hydrodynamic origin of quantum chaos in the presence of chiral anomaly. In  order to make it stronger, it would be nice if one shows that the analytic continuation of a hydrodynamic dispersion in the system passes through this point in the complex Fourier plane. We will discuss on this point in more details in \sec{conclusion}.




\section{Pole-skipping in the Green's function of  boundary operators}
\label{scalar_field}
As discussed in the Introduction, the chaos points are not the only well-known pole-skipping points in the Complex Fourier plane. In some specific cases, It has been shown that the pole of Green's functions of generic boundary operators skips at a set of points in the lower half of this plane. In this section we consider the near horizon dynamics of a scalar field perturbations on the magnetized black bane solution  \eqref{Eddington} to obtain such pole-skipping points. 

The scalar field equation of motion on the fixed background with metric $g_{\mu\nu}$ is given by
\begin{equation}\label{scalar_EoM}
\bigg(\partial_{\mu}(\sqrt{-g}\partial^{\mu})-m^2\sqrt{-g}\bigg)\Phi=0 .
\end{equation}
The strategy is as the following; one takes $\Phi=\phi(r)e^{-i \omega \nu+ i\vec{k}\cdot\vec{x}}$  with $\phi(r)$ expanded near the horizon as
\begin{equation}\label{phi_expanded}
\phi(r)=\sum_{n=0}^{\infty}\phi_{n}(r-r_h)^n=\,\phi_0+(r-r_h)\phi_1+\cdots.
\end{equation}
Then by putting \eqref{phi_expanded} back into \eqref{scalar_EoM},  the equation of scalar field can be rewritten in a near horizon expansion, giving a set of equations corresponding to different orders of expansion. The resultant equations are special in the sense that at  $\ell^{th}$ order  of expansion, one finds a linear equation between $\phi_0$, $\phi_1$, $\cdots$ and $\phi_{\ell+1}$, with the coefficient of $\phi_{\ell+1}$ vanishing at the Matsubara frequency $\omega_{\ell}=-i 2 \pi T (\ell+1)$.  In the following two subsections we will study the first four of these equations, namely those corresponding to $\ell=0,1,2,3$. They can be formally written as 
\begin{eqnarray}
0&=&M_{11}\phi_0+(2 \pi T-i \omega)\phi_1,\\
0&=&M_{21}\phi_0+M_{22}\phi_1+(4 \pi T-i \omega)\phi_2,\\
0&=&M_{31}\phi_0+M_{32}\phi_1+M_{33}\phi_2+(6 \pi T-i \omega)\phi_3,\\
0&=&M_{41}\phi_0+M_{42}\phi_1+M_{43}\phi_2+M_{44}\phi_3+(8 \pi T-i \omega)\phi_4
\end{eqnarray}
where $M_{rs}$ coefficients are in fact some functions of $\omega$  and $\vec{k}$.

As it is obvious from the above equations, just at the Matsubara frequency $\omega_{\ell}=-i 2 \pi T \ell$, the first $\ell$ equations decouple from the rest of them and take the following form
\begin{equation}
0	=\,\mathcal{M}_{\ell\times \ell}(\omega=-i2\pi T \ell, \vec{k})\begin{pmatrix}
\phi_0\\
\phi_1\\
.\\
.\\
\phi_{\ell-1}\\
\end{pmatrix}.
\end{equation} 
We already encountered with a similar situation when studying the near horizon dynamics  of $\delta g_{\nu\nu}$  perturbation at the chaos points in \eqref{delta_g_vv}. Similarly, the roots of equation $\det \mathcal{M}_{\ell\times \ell}(\omega=-i2\pi T \ell, \vec{k})=0$, here, are those wavenumbers at which, the ingoing boundary condition at the horizon is not sufficient to uniquely fix  a solution for $\Phi$ in the bulk, for a given UV normalization constant. Let us call the roots as $k_1, k_2, \cdots, k_{2\ell}$. It is clear that at these $2\ell$ points, the Green's function of the boundary operator, dual to the bulk scalar field $\Phi$, is multi-valued. 

According to the explanations given in the Introduction, based on \cite{Blake:2019otz}, one concludes that to every Matsubara frequency $\omega_{\ell}=-i 2\pi T \ell$, $\ell$ pole-skipping points of the dual boundary operator correspond. This simply shows that how near horizon dynamics can strictly constrain the Green's function of a generic boundary operator, even beyond the regime of hydrodynamics, namely at frequencies $\omega\sim T$. 

In the following two subsections, we find the set of pole-skipping points discussed above for the scalar field in the Einstein-Maxwell-Chern-Simons theory. Since there is a preferred direction, namely the magnetic field direction, which we take it as being along the third axis, our study of the scalar field perturbations falls into two cases; once when $\vec{k}\parallel \vec{B}$ and then when $\vec{k}\perp \vec{B}$ (see Appendix \ref{trans_poles} for the latter case).
\subsection{Longitudinal poles: a universal behavior}
Longitudinal pole-skipping points are obtained by studying the dynamics of the Fourier modes of $\Phi$ propagating in the direction of magnetic field. The Fourier components of the scalar field then are written as $\Phi=\phi(r)e^{-i \omega \nu+ ik x_3}$. Plugging in \eqref{scalar_EoM}, we arrive at
\begin{equation}
\begin{split}
&\frac{d}{dr}\left(\frac{e^{2 W_{T}(r)}\bigg((j(r)^2-e^{2 W_{L}(r)}F)\phi'(r)-i \omega (e^{2W_{L}(r)}q(r)-s(r) j(r))\phi(r)+ i k(F(r) s(r) - q(r) j(r) )\phi(r)\bigg)}{\sqrt{e^{2 W_{L}(r)}q(r)^2+ F(r) s(r)^2- 2 q(r) s(r) j(r)}}\right) \\
&+\frac{e^{2 W_{T}(r)}\bigg(\big(i k (F(r) s(r) - q(r) j(r))- i \omega (e^{2W_{L}(r)}q(r) -s(r)j(r))\big)\phi'(r)-( \omega s(r)+ k q(r))^2 \phi(r)\bigg)}{\sqrt{e^{2 W_{L}(r)}q(r)^2+ F(r) s(r)^2- 2 q(r) s(r) j(r)}}\\
&\\
&-e^{2 W_{T}(r)}\sqrt{e^{2 W_{L}(r)}q(r)^2+ F(r) s(r)^2- 2 q(r) s(r) j} \,\,m^2 \phi(r)=0
\end{split}
\end{equation}
As explained around \eqref{phi_expanded}, the above equation is equivalent to a set of linear equations for the near horizon components of  $\phi(r)$.
The first four equations are given by
\begin{eqnarray}\label{scalar_L_1}
0&=&M^{L}_{11}\phi_0+(2 \pi T-i \omega)\phi_1,\\\label{scalar_L_2}
0&=&M^{L}_{21}\phi_0+M^{L}_{22}\phi_1+(4 \pi T-i \omega)\phi_2,\\\label{scalar_L_3}
0&=&M^{L}_{31}\phi_0+M^{L}_{32}\phi_1+M^{L}_{33}\phi_2+(6 \pi T-i \omega)\phi_3,\\\label{scalar_L_4}
0&=&M^{L}_{41}\phi_0+M^{L}_{42}\phi_1+M^{L}_{43}\phi_2+M^{L}_{44}\phi_3+(8 \pi T-i \omega)\phi_4.
\end{eqnarray}
Here, the superscripts $L$ emphasize that $\vec{k}\parallel \vec{B}$. One can show that in this case,  all $M^{L}_{rs}$ coefficients can be generally written as the following:
\begin{equation} 
M^{L}_{rs}(\boldsymbol{\omega},\boldsymbol{k})=\, i \boldsymbol{\omega} \,a^{L}_{rs}+ \boldsymbol{k}^2 \,b^{L}_{rs}+ i \boldsymbol{\omega}\,\boldsymbol{k}\,c^{L}_{rs}+ d^{L}_{rs}
\end{equation}
where $\boldsymbol{\omega}=\omega/2\pi T$ and $\boldsymbol{k}=k/2 \pi T$. The coefficients $a_{rs}^L$, $b_{rs}^L$, $c_{rs}^L$ and $d_{rs}^L$ of the first three equations above have been given in the Table 2 of the Appendix \ref{Tables}.
\subsubsection{Pole-skipping points at $\omega_1=-i 2\pi T$}
From the near horizon equations of scalar field above, it is obvious that just at the lowest Matsubara frequency $\omega_{1}=-i 2\pi T$, equation \eqref{scalar_L_1}  decouples from the rest of equations. One then concludes that the pair of roots of $M_{11}^L\big(-i 2\pi T, i 2\pi T \boldsymbol{k}\big)=0$ correspond to the lowest frequency pole-skipping points. We call the roots as 
$\boldsymbol{k}^{L}_{1,j};\,\,j=1,2$ and find
\begin{equation}
\begin{split}
\boldsymbol{k}_{1,\{1,2\}}^{L}=&\pm\frac{1}{2}  \sqrt{m^2+6}\pm\frac{ \left(m^2+3\right) \nu ^2}{3 \sqrt{m^2+6}}-\kappa\,(\log (4)-1)\,  \nu ^2 b\mp\frac{ \left(\pi ^2 \left(m^2+6\right)-12 \left(m^2+3\right)\right)}{144 \sqrt{m^2+6}}\,b^2\\
\pm&\frac{ \left(18 \left(6 \left(24 \kappa ^2 (\log (2)-1)-11\right)+\pi ^2\right)+\left(\pi ^2-12\right) m^4+9 m^2 \left(16 \left(\kappa ^2 (\log (8)-3)-2\right)+\pi ^2\right)\right)}{216 \left(m^2+6\right)^{3/2}}\nu ^2b^2.
\end{split}
\end{equation}
\	
\begin{figure}[t]
	\centering
	\includegraphics[width=0.7\linewidth]{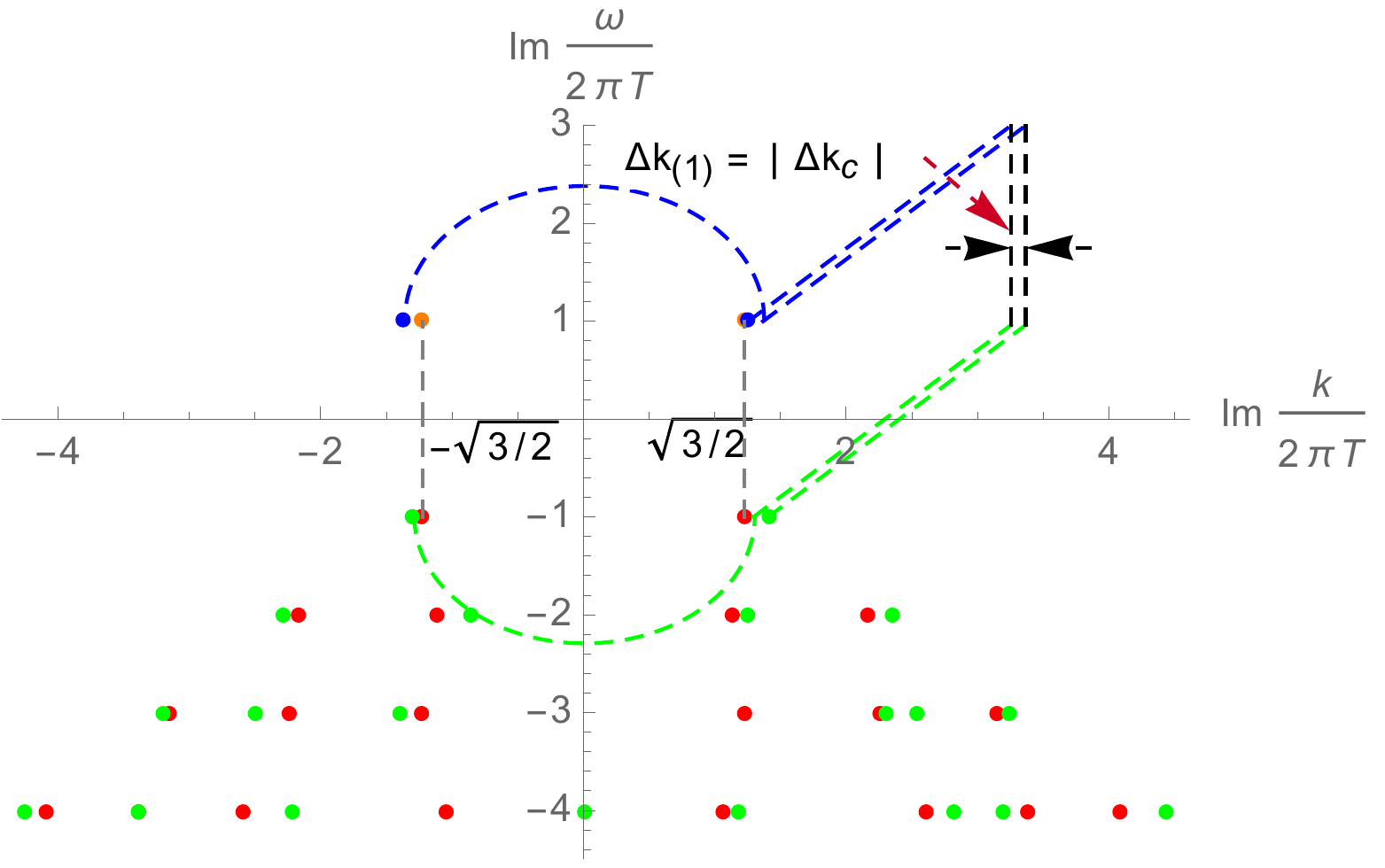}
	\caption{Spectrum of chaos points  (in the upper half plane) together with the pole-skipping points of a boundary  operator dual to a scalar field with mass $m=0$ in the bulk  (in the lower half plane). Orange and red points, which are symmetric with respect to the vertical axis, are related to non-chiral matter at $\nu=0$ and $b=0$. Blue and green points, however, correspond to a chiral matter at $\nu=0.5$ and $b=0.5$. Due to the chiral anomaly, these points are located asymmetrically with respect to vertical axis.   The vertical gray dashed lines at $\boldsymbol{k}=\pm\sqrt{3/2}$ show that at the special case when $m=0$, accidentally, the chaos points and the lowest frequency pole-skipping points have the same wavenumbers at $b=0$ and $\nu=0$.   }
	\label{chaos_pole_skippin_m_0}
\end{figure}
Obviously, these two points are not symmetric with respect to Im $\omega$-axis. The asymmetry is due to the chiral effects and, more precisely, is proportional with $\nu^2b$.  This is reminiscent of the asymmetry of the chaos points in \eqref{k_long_chaos_1_2}. The difference between the magnitude of $\boldsymbol{k}^{L}_{1,1}$ and $\boldsymbol{k}^{L}_{1,2}$ is
\begin{equation}\label{delta_k_1_MATSUBARA}
\Delta\boldsymbol{k}_{(1)}=\,\boldsymbol{k}_{1,1}^{L}-|\boldsymbol{k}_{1,2}^{L}|=\boldsymbol{k}_{1,1}^{L}+\boldsymbol{k}_{1,2}^{L}=\,-2 \,\kappa\,(\log (4)-1)\, \nu ^2b   
\end{equation}
	Interestingly, the term $\nu^2b$ in both $\boldsymbol{k}_{1,1}^L$ and $\boldsymbol{k}_{1,2}^L$ is independent of $m$, namely the mass of scalar field. Thus the above difference is \textit{universal} in the sense that it does not depend on the scaling dimension of the corresponding boundary operator \cite{Witten:1998qj}. 
	
\begin{figure}[t]
	\centering
	\includegraphics[width=0.7\linewidth]{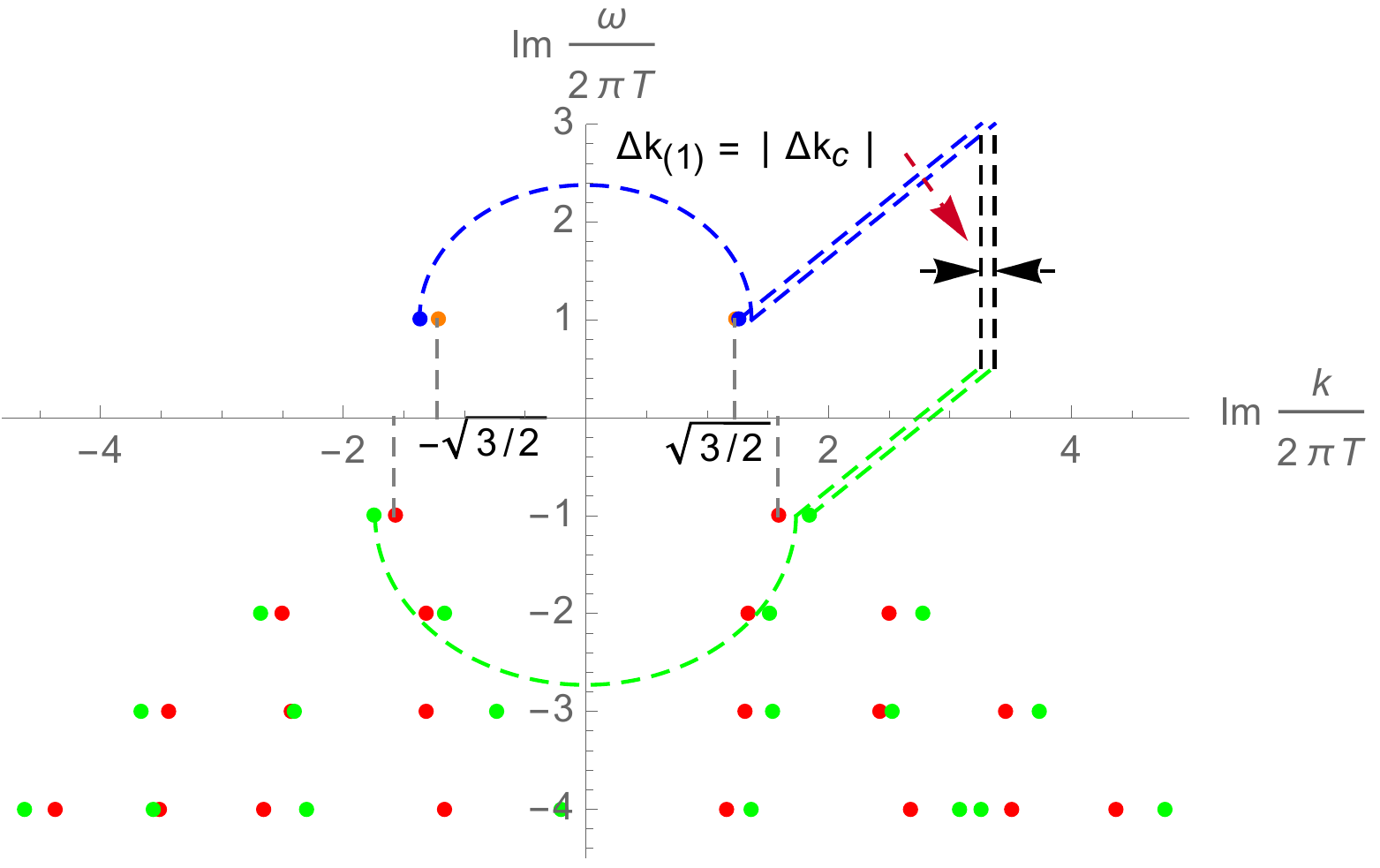}
	\caption{Spectrum of chaos points  (in the upper half plane) together with the pole-skipping points of a boundary  operator dual to a scalar filed with mass $m=2$ in the bulk  (in the lower half plane). Orange and red points are related to non-chiral matter at $\nu=0$ and $b=0$. Blue and green points correspond to a chiral matter at $\nu=0.5$ and $b=0.5$. Orange and red points are symmetric with respect to the Im
		$\omega$-axis, however, due to the chiral anomaly, the blue and Green ones are located asymmetrically with respect to the same axis.}
	\label{chaos_pole_skippin_m_2}
\end{figure}

	The universal behavior mentioned in the previous paragraph has been illustrated in both Figs.\ref{chaos_pole_skippin_m_0} and \ref{chaos_pole_skippin_m_2}. 
In these figures, we have also demonstrated the chaos points found in \eqref{k_long_chaos_1_2}. 
As it can be seen, in general, there is no explicit relation between the position of chaos points (the blue ones) and that of the lowest frequency pole-skipping points (the highest pair of green ones). However, there is an implicit relation between them; \textit{the sum of wavenumbers corresponding to pole-skipping point with the lowest Matsubara frequency, in the lower half plane, is exactly the absolute value of that of chaos points, given by \eqref{k_chaos}, in the upper half plane}: $\Delta\boldsymbol{k}_{(1)}=|\Delta \boldsymbol{k}_{c}|$. We will comment on this interesting property later.
\subsubsection{Pole-skipping points at $\omega_2=-i 4\pi T$ and higher Matsubara frequencies}
At the next Matsubara frequency, namely at $\omega_2=-i 4\pi T$, two of the near horizon equations (\eqref{scalar_L_1} and \eqref{scalar_L_2}) decouple form the rest of them.
The corresponding pole-skipping points can then be found by studying $M^L_{rs}$ coefficients of the first two equations. Solving 
\begin{equation}
\begin{split}
0&=det \mathcal{M}_{2\times2}(-i 4 \pi T, i 2\pi T \boldsymbol{k})\\
&=M_{11}^{L}(-i4\pi T,i2\pi T \boldsymbol{k})M_{22}^{L}(-i4\pi T,i2\pi T \boldsymbol{k})+(2\pi T)M_{12}^{L}(-i4\pi T,i2\pi T \boldsymbol{k}),
\end{split}
\end{equation}
one finds four roots  $\boldsymbol{k}^{L}_{2,4}<\boldsymbol{k}^{L}_{2,3}<\boldsymbol{k}^{L}_{2,2}<\boldsymbol{k}^{L}_{2,1}$, corresponding to two pairs of pole skipping points, namely $\{\boldsymbol{k}^{L}_{2,1}, \boldsymbol{k}^{L}_{2,4}\}$ and $\{\boldsymbol{k}^{L}_{2,2}, \boldsymbol{k}^{L}_{2,3}\}$. In the absence of chiral effects, these four are represented by the red points at Im$\frac{\omega}{2\pi T}=\,-2$  in Figs.\ref{chaos_pole_skippin_m_0} and \ref{chaos_pole_skippin_m_2}. It is shown that the wave number of red points satisfy the following relations \cite{Blake:2019otz}
\begin{equation}
\boldsymbol{k}^{L}_{2,1}=-\boldsymbol{k}^{L}_{2,4},\,\,\,\,\,\,\,\,\,\boldsymbol{k}^{L}_{2,2}=-\boldsymbol{k}^{L}_{2,3}.
\end{equation} 
At the same row of the figures, however, there are four green points showing the pole-skipping
in our holographic chiral system; one obviously sees that the above equalities will no longer hold for the green points.
Although the analytic expressions of the wavenumbers for the green points are so much complicated,  their deviation from the symmetric case can be written in a simple form:
\begin{eqnarray}
\boldsymbol{k}^{L}_{2,1}+\boldsymbol{k}^{L}_{2,4}&=&-4 \kappa  \left(\log (4)-1-\frac{1}{\sqrt{2m^2+12}}\right) \nu ^2b \\
\boldsymbol{k}^{L}_{2,2}+\boldsymbol{k}^{L}_{2,3}&=&-4   \kappa   \left(\log (4)-1+\frac{1}{\sqrt{2m^2+12}}\right) \nu ^2b.
\end{eqnarray}
 Compared to the asymmetry of the lowest frequency points given by \eqref{delta_k_1_MATSUBARA},  it seems that these results are not universal, in the sense  that they explicitly depend on the mass of scalar field. However, the sum of all four wavenumbers is interestingly mass independent:
 \begin{equation}\label{delta_k_2_MATSUBARA}
 	\Delta\boldsymbol{k}_{(2)}=\,\boldsymbol{k}_{2,1}^{L}+\boldsymbol{k}_{2,2}^{L}+\boldsymbol{k}_{2,3}^{L}+\boldsymbol{k}_{2,4}^{L}=\,-8 \,\kappa\,(\log (4)-1)\, \nu ^2b.
 \end{equation}
  It suggests that perhaps \textit{the sum of wavenumbers of pole-skipping points at a specific Matsubara frequency encodes some information about chiral anomaly}. To explore more on this idea, we compute
 \begin{equation}\label{univ_quantity}
 \Delta \boldsymbol{k}_{(\ell)}=\,\sum_{r=1}^{2\ell}\boldsymbol{k}^L_{\ell,r}\,\,\,\,\,\,\,\,\,\,\,(\omega_{\ell}=-i 2\pi T \ell)
 \end{equation}
 for $\ell=1,2,3,4$. In this formula, $\boldsymbol{k}^L_{\ell,r}$ is the wave number corresponding to the $r^{th}$ longitudinal pole-skipping point at $\ell^{th}$ Matsubara frequency. Surprisingly, we obtain
 \begin{eqnarray}
 \Delta \boldsymbol{k}_{(1)}&=&-2 \,\kappa\,(\log (4)-1)\,    \nu ^2 b\\
  \Delta \boldsymbol{k}_{(2)}&=&-8\,\kappa\,(\log (4)-1)\,   \nu ^2 b\\
   \Delta \boldsymbol{k}_{(3)}&=&-18\,\kappa\,(\log (4)-1)\,  \nu ^2 b\\
    \Delta \boldsymbol{k}_{(4)}&=&-32 \,\kappa\,(\log (4)-1)\,   \nu ^2 b.
 	\end{eqnarray}
 	Considering the above expressions, one may conjecture that the sum of wavenumbers corresponding to the pole-skipping points at $\ell^{th}$ Matsubara frequency $ 	\omega_{\ell}=-i 2\pi T$, is given by the following closed formula
 	\begin{equation}\label{closed}
 \boxed{\Delta\boldsymbol{k}_{(\ell)}=\,-2\ell^2\, \kappa\, \big(\log(4)-1\big)\,\nu^2b}
 	\end{equation}
 	This  results is not only surprising because of its simple feature, but is also important from the viewpoint of its universality. It can be seen that whatever the mass of scalar field in the bulk and correspondingly the scaling dimension of the dual boundary operator is,  this result always identically holds. In the following subsection, we discuss about the relation between \eqref{closed} and quantum chaos.
 	\subsection{Longitudinal poles and quantum chaos}
 What we obtained in previous subsection can be reviewed as it follows. Firstly, it is important to note that due to the chiral effects, the arrangement of green pole-skipping points in Fig.\ref{chaos_pole_skippin_m_0} and Fig.\ref{chaos_pole_skippin_m_2} is asymmetric with respect to Im $\omega$ axis. We showed that an appropriate quantity which universally captures the asymmetric feature is the one introduced in \eqref{univ_quantity}. We finally argued that this quantity may in general follow from a closed formula given by \eqref{closed}. Comparing this result with \eqref{k_chaos}, we arrive at
 	\begin{equation}\label{last_result}
 	\frac{\Delta\boldsymbol{k}_{(\ell)}}{\Delta\boldsymbol{k}_{c}}=-\ell^2
 	\end{equation}
 One concludes that in addition to the splitting of butterfly velocities, the non-zero value of the quantity $\Delta \boldsymbol{k}_{(\ell)}$ can be regarded as the macroscopic manifestation of chiral anomaly. 
 It would be also interesting to investigate whether the above relation continues to hold beyond the perturbative limit of small $\nu$ and $b$. We leave more investigation about it to a future work. 
\section{Conclusion and outlook}
\label{conclusion}
In this paper we holographically studied  quantum chaos and pole-skipping phenomenon in a system with chiral anomaly. Using the shock wave picture, we first computed the butterfly velocities in both directions parallel and perpendicular to the magnetic field. 
As we showed, in the parallel case, the two butterfly velocities would split due to the chiral anomaly. Then by computing the spectrum of hydrodynamic modes in the system, we found that the velocity of sound waves propagating parallel and anti-parallel to the magnetic field are split for the same reason. To leading order in small $B/T^2$ and $\mu/T$, we arrived at \eqref{chaos_hydro}. It would be interesting to investigate whether  this relation continues to hold beyond the perturbation regime of $\nu$ and $b$.\footnote{If it is the case, then following \cite{Grozdanov:2016ala}, it would be interesting to investigate whether such feature will be protected at finite coupling. }
 
 The splitting of butterfly velocities has an important implication. Let us recall that the butterfly velocity can be actually measured in the experiment \cite{Swingle:2016var,Danshita:2016xbo,Garttner:2016mqj,Li:2017pbq}. The splitting of butterfly velocities due to chiral anomaly then implies that designing relevant experiments to measure OTOC's in chiral matters, e.g. in the Weyl semimetal \cite{Yan:2017jgt,Burkov:2017rgl}, might be important for macroscopically detecting the chiral anomaly. 

To investigate the relation between quantum chaos and hydrodynamics, we then holographically computed the pole-skipping points in the energy density Green's function of the boundary theory. The pole-skipping points were found to be precisely the same as chaos ones \eqref{coinsidence}. This coincidence might be regarded as hydrodynamic origin of quantum chaos in an anomalous system. 
However, there are yet another ways to explore the relation between hydrodynamics and quantum chaos. Following \cite{Grozdanov:2017ajz} and in the language of \cite{Grozdanov:2019uhi}, one can investigate whether the chaos points given in \eqref{k_long_chaos_1_2} are located on the analytically continued dispersion relation of sound waves. To proceed, one has to firstly find the spectral curve corresponding to the sound channel.   To this end, it is needed to know the gauge invariant objects in this channel. One can show that for the general class of the solutions 
\begin{eqnarray*}
ds^2&=&-a(r) dt^2+b(r)dr^2+a_1(r)(dx_1^2+dx_2^2)+a_3(r)dx_3^2+2 c(r)dt dx_3\\
A&=&\left(\int E(r)dr\right) dt-\frac{B}{2}\big(x_2 dx_1-x_1 dx_2\big) -\left(\int P(r)dr\right) dx_3,
\end{eqnarray*}
in the sound channel, only 
$h_{tt}$, $h_{x^3x^3}$, $h_{x^ix^i}$, $h_{rr}$, $h_{rt}$, $A_t$, $A_r$ and $A_{x^3}$ mix with each other.
We have assumed metric and gauge field perturbations as   $h_{\mu \nu}e^{-i \omega t+i k x_3}$ and $A_{\mu }e^{-i \omega t+i k x_3}$, respectively.  It can then be shown that there are two relevant gauge invariant quantities constructed out of the metric and gauge perturbations in this channel:
\begin{eqnarray*}
E_{z}&=& k A_t+ \omega A_{x^3}+\frac{k E(r)-\omega P(r)}{a_1'(r)}h_{x^i x^i}\\
Z&=&k^2 h_{tt}+\omega^2h_{x^3 x^3}+2\omega k\, h_{tx^3}+\frac{k^2a'(r)-\omega^2a_{3}'(r)-2 k \omega \,c'(r)}{ a_1'(r)}h.
\end{eqnarray*}
Finally it turns out that these two quantities obey two coupled second order ordinary differential equations. Formally one writes
\begin{eqnarray*}\label{Sound_channel_diff}
Z''+\,{{\textswab{a}}}_{1}\,Z'+\,{{\textswab{a}}}_{2}\,Z+\,{{\textswab{a}}}_{3}\,E_z+\,{{\textswab{a}}}_{4}\,E_z'&=&0\\
E_z''+\,{{\textswab{b}}}_{1}\,E_z'+\,{{\textswab{b}}}_{2}\,E_z+\,{{\textswab{b}}}_{3}\,Z+\,{{\textswab{b}}}_{4}\,Z'&=&0
\end{eqnarray*}
with the coefficients being some functions of $\omega$, $k$ as well as metric and gauge field components. Considering the radial coordinate in the bulk as $u\sim1/r^2$, one then looks for solutions $Z(u;\omega,k^2,kB)$ and $E_z(u;\omega,k^2,kB)$ to the above equations. Solutions must be regular on the future event horizon and also obey the Dirichlet boundary condition at $u=0$.  The latter gives spectral curve of sound channel $Z(0;\omega,k^2,kB)=0$ and $E_z(0;\omega,k^2,kB)=0$.
Then by using the method developed in \cite{Grozdanov:2019uhi}, one can find the dispersion relation of sound modes in an expansion over both $k$ and $B$. If it behaves like the non-chiral case \cite{Grozdanov:2017ajz}, luckily, the first few orders in the expansion will be sufficient to observe that chaos point is on the analytic continuation of this spectral curve. If so, it would be also interesting to investigate whether chaos points are located within the domain of convergence of hydrodynamics in our system or not.

In another direction, we can fully do the above task numerically. One can numerically solve the above coupled differential equations for a given specific wave number to find the corresponding spectrum of quasi normal modes in the system. However, for purely
imaginary values of the wave number, in addition to the points with negative imaginary
frequency, one finds a point in the upper half of the complex plane, just on vertical axis.
Following the effective field theory discussion of \cite{Blake:2017ris}, one can collect a number of these points and interpolate among
them to see whether the chaos points are on the interpolating curve.
We leave detailed computations to a future work. 

In the last part of the paper, we showed that the same information about splitting the butterfly velocities is encoded in pole-skipping points of Green's function associated with a boundary operator dual to the scalar field in the bulk. More precisely, we firstly computed the spectrum of pole-skipping points in the lower half of complex Fourier plane. Then we showed that the sum of wavenumbers corresponding to pole-skipping points at a specific Matsubara frequency was a universal quantity. By universal we mean that the sum does not depend on the mass of scalar field in the bulk and consequently neither does on the scaling dimension of dual boundary operator. 
More interestingly, we proposed a closed formula for the above mentioned universal quantity, in \eqref{closed}.
While formula \eqref{closed} was found through holographic computations, it would be interesting to investigate the reason behind this nice behavior of pole-kipping points, in quantum field theory. 

\section*{Acknowledgment}
We would like to thank Ali Davody for valuable discussions on our results. We also thank Armin Ghazi,  Karl Landsteiner, Shu Lin and Omid Tavakol for discussion.
 We are particularly grateful to Richard Davison and Saso Grozdanov for reading the draft and giving many helpful comments.
\appendix
\section{Detailed computations of the longitudinal butterfly velocity}
\label{App_contour}
\begin{figure}
	\centering
	\begin{tikzpicture}
	[decoration={markings,
		mark=at position 3cm with {\arrow[line width=1pt]{>}},
		mark=at position 7cm with {\arrow[line width=1pt]{>}},
		mark=at position 8.9cm with {\arrow[line width=1pt]{>}},
		mark=at position 15cm with {\arrow[line width=1pt]{>}}
	}
	]
	\draw[help lines,->] (-4,0) -- (4,0) coordinate (xaxis);
	\draw[help lines,->] (0,-4) -- (0,4) coordinate (yaxis);

	
%
	\node at (.5,1.2) {$im_0$};
		\node at (-.7,-1.2) {$-im_0$};
						\node at (0,1.2) {$\bullet$};
		\node at (0,-1.2) {$\bullet$};
				\node[blue] at (0,1.9) {$\bullet$};
						\node[red] at (0,-.5) {$\bullet$};
								\node[blue] at (-1.6,1.9) {$i(p+\sqrt{p^2+m_0^2})$};
										\node[red] at (1.7,-.5) {$i(p-\sqrt{p^2+m_0^2})$};
	
	
	
						\path[dashed,draw,blue,line width=0.8pt,postaction=decorate] (-3.,.1) -- (3.,.1) coordinate ;
									\path[dashed,red,draw,line width=0.8pt,postaction=decorate] (-3.,-.1) -- (3.,-.1) coordinate ;
																	\path[dashed,draw,red,line width=0.8pt,postaction=decorate] (3,-.1) arc (0:-180:3);
																	\path[dashed,blue,draw,line width=0.8pt,postaction=decorate] (3,0.1) arc (0:180:3);
														\node at (2.8,2.4) {$\boldsymbol{C_{1}}$};
																	\node at (2.8,-2.4) {$\boldsymbol{C_{2}}$};
																		\path[dashed,red,draw,line width=0.99pt,postaction=decorate,->]  (0,-1.2) -- (0,-.6) coordinate ;
																			\path[dashed,blue,draw,line width=0.99pt,postaction=decorate,->]  (0,1.2) -- (0,1.8) coordinate ;
																			
	\node[above] at (xaxis) {$\text{Re} k$};
	\node[left] at (yaxis) {$\text{Im} k$};
	\end{tikzpicture}
	\caption{Asymmetric analytic structure of $\tilde{h}$ in the presence of magnetic field and chiral anomaly. Black points, which are symmetric with respect to the real axis, show the location of poles in the absence of magnetic filed.}
	\label{complex_h_3}
\end{figure}
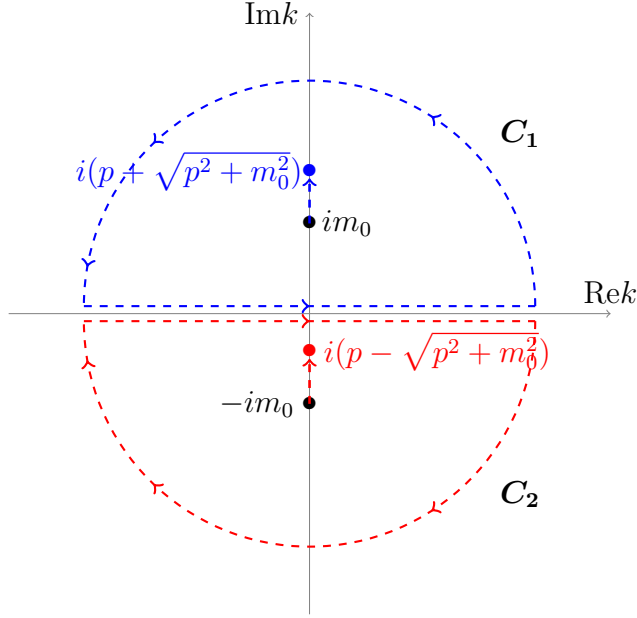

When $h$ is just a function of $x_3$, equation \eqref{formal_shock_equation} takes the following form
\begin{equation}\label{formal_shock_equation_L}
\left( \partial_{\parallel}^2 +2 \,{\textswab{p}}\, \vec{b}\cdot \vec{\partial}- m_0^2\right)h(x)\sim \frac{2 B_{L}(0)}{A(0)} E e^{\frac{1}{2}\tilde{f}'(r_h)t}\,\delta(x_3).
\end{equation}
As before, we assume the magnetic field to be directed along $+x_3$ direction. In the Fourier space, $h(x_3)= \tilde{h} \,e^{i k\, x_3}$ and one finds
\begin{equation}\label{formal_shock_equation_L_Fourier}
\left( -k^2 +2 i\,{\textswab{p}}\, b\, k- m_0^2\right)\tilde{h}\sim \tilde{E}
\end{equation}
with $\tilde{E}=\frac{2 B_{L}(0)}{A(0)} E e^{\frac{1}{2}\tilde{f}'(r_h)t}$. As a result, the function $h$ can be found by performing the following integral: 
\begin{equation}\label{}
h(x_3)\sim \int_{-\infty}^{+\infty} d k\, e^{i k x_3}\frac{-\tilde{E}}{k^2- 2 i {\textswab{p}} \,k+m_0^2}=\int_{-\infty}^{+\infty} dk e^{i k x_3}\frac{-\tilde{E}}{(k-k_1)(k-k_2)}
\end{equation}
with $k_1= i({\textswab{p}}+\sqrt{{\textswab{p}}^2+m_0^2})$ and $k_2= i({\textswab{p}}-\sqrt{{\textswab{p}}^2+m_0^2})$.  To proceed, we close a contour in the complex plane of $k$ (See Fig. \ref{complex_h_3}).
In the absence of magnetic field, $h(x_3)$ is symmetric under $x_3 \leftrightarrow -x_3$. However, a magnetic field directed in $+x_3$ direction breaks down this symmetry. When $x_3>0$, the integral along $C_1$ is convergent, while for $x_3<0$ integration over $C_2$ converges. One then finds
\begin{equation}
h(x_3)\sim (2\pi i)(-\tilde{E})\left[\frac{e^{-({\textswab{p}}+\sqrt{{\textswab{p}}^2+m_0^2})x_3}}{2 i \sqrt{{\textswab{p}}^2+m^2}}\theta(x_3)+\frac{e^{-({\textswab{p}}-\sqrt{{\textswab{p}}^2+m_0^2})x_3}}{2 i \sqrt{{\textswab{p}}^2+m^2}}\theta(-x_3)\right].
\end{equation}
Using this, one simply obtains \eqref{butterfly_from_shcok}. In fact, the asymmetric positioning of the poles with respect to the real axis in Figure.\ref{complex_h_3} is the origin of splitting between the butterfly velocities.

\section{Transverse butterfly and pole-skipping}
\label{App_transverse}
If function $h$ in \eqref{formal_shock_equation}  is assumed to be a function of the transverse directions, say $x_{\perp}=x_1,x_2$, then this equation simplifies to 
\begin{equation}\label{formal_shock_equation_T}
\left(\frac{}{} \partial _{\perp}^2 - \frac{m_0^2}{{\textswab{q}}^2}\right)h(x_{\perp})\sim \frac{2 B_{L}(0)}{A(0){\textswab{q}}^2} E e^{\frac{1}{2}\tilde{f}'(r_h)t}\,\delta^2(x_\perp).
\end{equation}
It has the following simple solution
\begin{equation}
h(x_{\perp})\sim \,E \,e^{\frac{1}{2}\tilde{f}'(r_h)(t_\omega-t_*)-\frac{m_0}{{\textswab{q}}}|x_{\perp}| } g(|x_{\perp}|)
\end{equation}
where $g$ is a non-exponential function. 
The transverse speed of propagation then reads
\begin{equation}
v_{B}^{T_{1,2}}=\,\frac{2\pi T}{m_0}\,{\textswab{q}}.
\end{equation}
Using ${\textswab{q}}=1-\frac{\pi^2}{24}b^2$ together with \eqref{p_m_o}, one simply finds 
\begin{equation}\nonumber
v_{B}^{T_{1,2}}=\pm\sqrt{\frac{2}{3}}\left(1-\frac{\mu^2}{3(\pi T)^2}\right)\pm \left(-\frac{12+\pi ^2}{72 \sqrt{6}}+\frac{ \left(\pi ^2+36 \left(4 \kappa ^2 (\log (4)-2)-3\right)\right)}{216 \sqrt{6}}\frac{\mu^2}{(\pi T)^2}\right)\frac{B^2}{(\pi T)^4}
\end{equation}
There are two butterfly velocities with the same magnitude in the transverse directions. Although the magnitude of the butterfly velocity has changed compared to the non-chiral system, however, there is no any term linearly depending on the magnetic field. In other words, the magnetic field starts to contribute like $\frac{\mu^2B^2}{T^6}$ for both transverse butterfly velocities. This basically means that in this case it is just by precisely measuring the butterfly velocities that one can explore the existence of chiral effects. Let us recall that in the longitudinal case, any difference between the magnitude of the two butterfly velocities was regraded as the sign of such anomalous effects.

In the absence of chiral effects, namely when $\kappa=0$, the above result simplifies to
\begin{equation}\label{non_chiral_v_B_Trans}
v_{B, \kappa=0}^{T_{1,2}}=\pm v_{B}\left[1-\frac{\mu^2}{3(\pi T)^2}-\left(\frac{1}{12}+\frac{\pi^2}{144}\right)\frac{\mu^2}{(\pi T)^2}+\left(-\frac{1}{4}+\frac{\pi^2}{432}\right)\frac{\mu^2B^2}{(\pi T)^6}\right].
\end{equation}
We discuss more about this equation in the following subsection.

Now we show that the transverse pole-skipping points of the energy density Green's function precisely coincide with the transverse chaos points $(\omega_{c}, k_{c}^{T})=(i 2\pi T ,i 2 \pi T/v_{B}^{T_{1,2}})$.  
When $\vec{k}\perp \vec{B}$, the perturbations $\delta g_{vv}$, $\delta g_{rr}$, $\delta g_{rv}$, $\delta g_{x^ix^i}$, $\delta g_{x^3x^3}$, $\delta g_{vx^1}$, $\delta g_{vx^2}$,  $\delta g_{rx^1}$ and $\delta g_{rx^2}$ decouple form the others. Considering these metric perturbations around \eqref{Eddington}, the linearized $vv$ component of Einstein equations near the horizon then, up to second order in $b$ and $\nu$, reads:
\begin{equation}\label{NH_linear_Trans}
\begin{split}
&\bigg[	k^2-3i \pi T \omega+\left(-\frac{4}{3}k^2+2i \pi T \omega\right)\nu^2+\left(-\big(\frac{\pi^2}{72}+\frac{1}{3}\big) k^2+\frac{1}{2}i \pi T \omega\right)b^2+\mathcal{C}\,\nu^2b^2\bigg]\,\,\delta g_{vv}^{(0)}\\
&	-i (2\pi T+i \omega)\frac{4\nu^2-3}{216}\bigg[ (-72+(\pi^2+24)b^2)\bigg(2k\,\delta g_{vx^3}^{(0)}+\omega \delta g_{x^ix^i}^{(0)}\bigg)-2\omega\left(36+(\pi^2-12)b^2\right)\delta g_{x^3x^3}^{(0)}\bigg]=\,0
\end{split}
\end{equation}
with the coefficients $\mathcal{C}=2 i \pi T \,\bigg[(\log(4)-2)\kappa^2-1\bigg]\, \omega\, \nu^2 b^2+\frac{24+\pi^2}{54}k^2\, \nu^2 b^2$. 
 Again, like the longitudinal case, at the frequency $\omega=i 2\pi T$, the perturbation $\delta g_{vv}$ decouples from the other perturbations at the horizon. We find the position of pole-skipping points $(\omega^{*} , k^{*})=(i 2 \pi T,i 2 \pi T \boldsymbol{k}^*)$ with $\boldsymbol{k}^*$ given by
 \begin{equation}
\boldsymbol{k}_{1,2}^*= \pm\sqrt{\frac{3}{2}}\pm\frac{\nu ^2}{\sqrt{6}}\pm \left(\frac{12+\pi ^2}{48 \sqrt{6}}+\frac{ \left(-144 \kappa ^2 (\log (4)-2)+\pi ^2+132\right)}{144 \sqrt{6}}\nu ^2\right)b^2.
 \end{equation}
One can simply check that two values of $\boldsymbol{\omega}^*/\boldsymbol{k}^{*}$ in this case are exactly equal to the butterfly velocities \eqref{non_chiral_v_B_Trans}. This means that in an anomalous system, the pole-skipping points are the same as chaos points, not only in the longitudinal case, but also in the transverse directions. Thus it completes our discussions about the relation between hydrodynamics and quantum chaos in an anomalous system. 
\section{Comparison with Blake-Davison-Sachdev\cite{Blake:2017qgd} }
\label{Blake_Davison_Sachdev}
In \cite{Blake:2017qgd}, based on the shock wave propagation picture in the bulk of AdS \cite{Shenker:2013pqb},  the butterfly velocities for an anisotropic Q-lattice has been computed. Considering the solution as 
\begin{equation}\label{Blake_metric}
ds^2=\,-f(r)dt^2+\frac{dr^2}{f(r)}+h_{T}(r)\,(dx_1^2+dx_2^2)+h_{L}(r) \,dx_3^2,
\end{equation}
authors of \cite{Blake:2017qgd} show that butterfly velocities in the longitudinal ($L$) and transverses ($T$) directions  are given as the following 
 \begin{equation}\label{Blake_butterfly}
v_{L}=\frac{2 \pi T}{\sqrt{h_L} m}\bigg|_{r_h},\,\,\,\,\,\,\,\,v_{T}=\frac{2 \pi T}{\sqrt{h_T} m}\bigg|_{r_h},\,\,\,\,\,\,\,m^2=\,\pi T\left(\frac{2h'_T h_L+h'_l h_T}{h_Th_L}\right)\bigg|_{r_h}.
 \end{equation}
 with $m$ being the effective mass of the shock wave.
 Let us recall that in the current paper, we have been mainly working on the solution \eqref{metric}.
 Obviously, when $C(r)=0$, namely when the solution is non-chiral but still magnetized, \eqref{metric} reduces to \eqref{Blake_metric}, with the following identifications:
\begin{equation}
h_{T}(r)\equiv\,e^{2 W_{T}(r)},\,\,\,\,\,\,\,\,\,h_{L}(r)\equiv e^{2 W_{L}(r)}.
\end{equation}
The expressions of $W_T(r)$ and $W_L(r)$ have been given in the Table 1 of the Appendix \ref{Tables}. Using them, we find 
\begin{eqnarray}
m|_{r_h}&=&\sqrt{6}\left(1-\frac{\nu^2}{3}\right)+\frac{1}{2\sqrt{6}}\left(-1+\frac{11}{3}\nu^2\right)b^2\\
h_{T}(r_h)&=&\pi^2 T^2\left[1+\frac{4}{3}\nu^2-\frac{\pi^2-12}{9}\left(\frac{1}{4}+\frac{\nu^2}{3}\right)b^2\right]\\
h_{L}(r_h)&=&\pi^2 T^2\left[1+\frac{4}{3}\nu^2+\frac{\pi^2+24}{18}\left(\frac{1}{4}+\frac{\nu^2}{3}\right)b^2\right].
\end{eqnarray}
By use of the recent expressions, then, the longitudinal and transverse butterfly velocities given in \eqref{Blake_butterfly} turn out to be exactly the same as \eqref{v_long_non_chiral} and \eqref{non_chiral_v_B_Trans}, respectively.

\section{Thermodynamic derivatives}
\label{thermo_derivatives}
Using the equation of sate given in \eqref{delta_epsilon}, \eqref{delta_pressure} and \eqref{delta_charge}, one finds
\begin{eqnarray}
\alpha_1&=&\left(\frac{\partial \epsilon}{\partial T}\right)_{\mu}=\,N_c^2 \pi^2 T^3\left[\frac{3}{2}+3 \nu ^2+b^2 \left(\nu ^2 \left(\frac{8}{3} \pi  \log (2)-\pi \right)-\frac{1}{4}\right)\right]\\
\alpha_2&=&\left(\frac{\partial \epsilon}{\partial \mu}\right)_{T}=\,N_c^2 \pi T^3\,\nu\left[3+b^2   \left(\pi -\frac{8}{3} \pi  \log (2)\right)\right]\\
\beta_1&=&\left(\frac{\partial n}{\partial T}\right)_{\mu}=\,N_c^2\pi T^2\nu\left[2 + b^2   \left(2-\frac{16 \log (2)}{3}\right)\right]\\
\beta_2&=&\left(\frac{\partial n}{\partial \mu}\right)_{T}=\,N_c^2 T^2 \left[1+4 \nu ^2+b^2 \left(\frac{8 \log (2)}{3}-1\right)\right]\\
\gamma_1&=&\left(\frac{\partial p}{\partial T}\right)_{\mu}=\,N_c^2 \pi^2 T^3\left[\frac{1}{2}+\nu ^2+b^2 \left(\nu ^2 \left(\pi -\frac{8}{3} \pi  \log (2)\right)+\frac{1}{4}\right)\right]\\
\gamma_2&=&\left(\frac{\partial p}{\partial \mu}\right)_{T}=\,N_c^2 \pi T^3\nu\left[1+b^2   \left(\frac{8}{3} \pi  \log (2)-\pi \right)\right]
\end{eqnarray}
\section{Scalar field dynamics in transverse directions: symmetric spectrum}
\label{trans_poles}
Transverse pole-skipping points are obtained by studying the dynamics of Fourier modes of $\Phi$ propagating in the directions perpendicular to the magnetic field. As before, we take the magnetic field along the third axis. Exploiting $SO(2)$ rotational symmetry in the transverse plane, we then take the Fourier components of the scalar field as $\Phi=\phi(r)e^{-i \omega \nu+ ik x_1}$. Plugging in \eqref{scalar_EoM}, we arrive at
\begin{equation}
\begin{split}
\frac{d}{dr}&\left(\frac{e^{2 W_{T}(r)}\bigg((j(r)^2-e^{2 W_{L}(r)}F)\phi'(r)-i \omega (e^{2W_{L}(r)}q(r)-s(r) j(r))\phi(r)\bigg)}{\sqrt{e^{2 W_{L}(r)}q(r)^2+ F(r) s(r)^2- 2 q(r) s(r) j(r)}}\right) \\
&\,\,\,\,\,\,\,\,\,\,\,+\frac{e^{2 W_{T}(r)}\bigg(\big(- i \omega) (e^{2W_{L}(r)}q(r)-j(r)s(r))\big)\phi'(r)- \omega^2s^2(r) \phi(r)\bigg)}{\sqrt{e^{2 W_{L}(r)}q(r)^2+ F(r) s(r)^2- 2 q(r) s(r) j(r)}}\\
&\\
&\,\,\,\,\,\,\,\,\,\,\,\,\,\,\,\,\,\,\,\,\,-e^{2 W_{T}(r)}\sqrt{e^{2 W_{L}(r)}u(r)^2+ F(r) s(r)^2- 2 q(r) s(r) j} \,\,m^2 \phi(r)=0.
\end{split}
\end{equation}
\begin{figure}[t]
	\centering
	\includegraphics[width=0.48\linewidth]{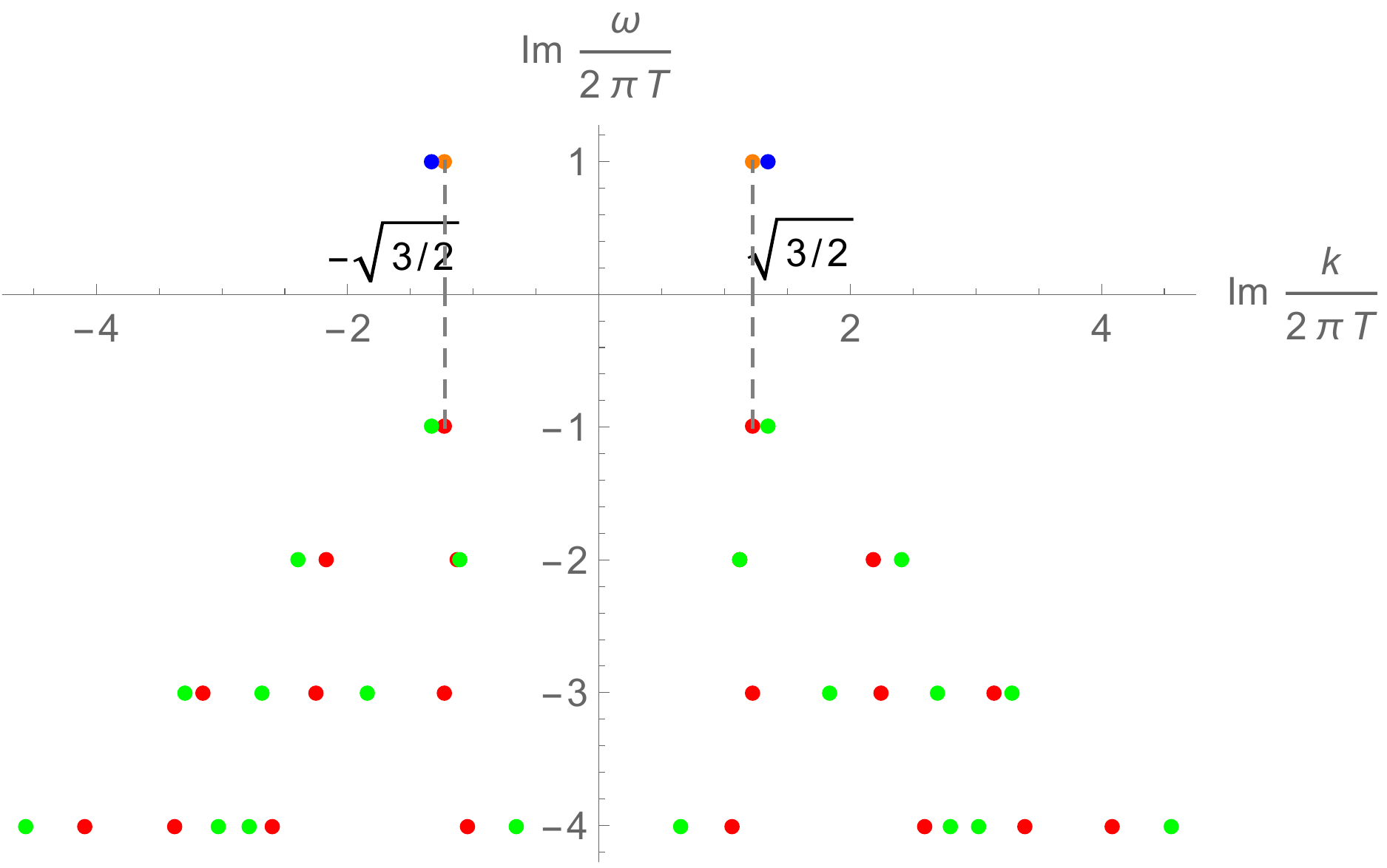}
	\includegraphics[width=0.48\linewidth]{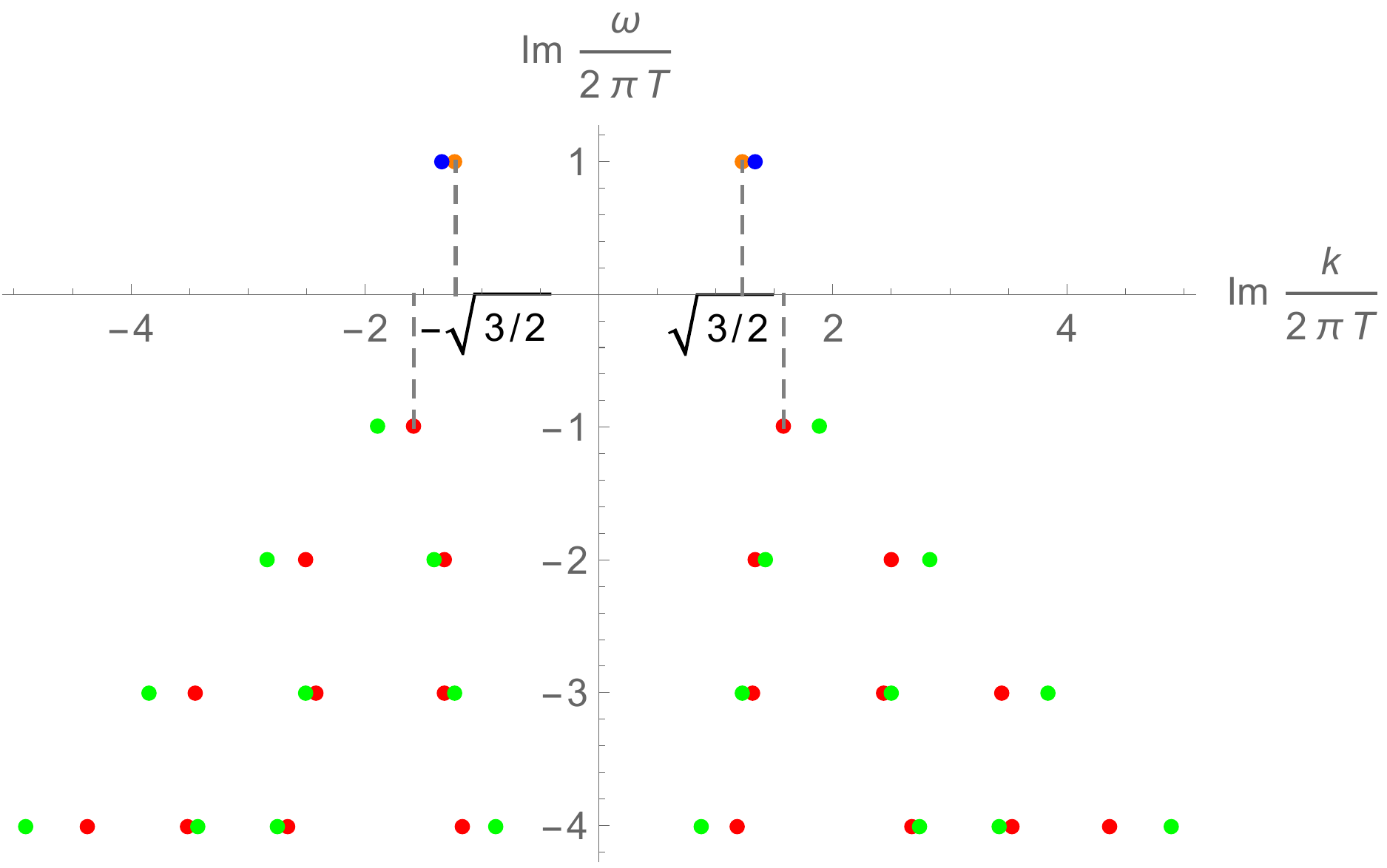}
	\caption{Spectrum of chaos points together with the pole-skipping points of a boundary  operator dual to a scalar filed, in the \textit{transverse} channel. In the left panel, the mass of scalar field in the bulk is $m=0$ and in the right panel $m=2$. Orange and red points are related to non-chiral matter at $\nu=0$ and $b=0$. Blue and green points correspond to a chiral matter at $\nu=.5$ and $b=.5$.}
	\label{chaos_pole_skippin_T_m_02}
\end{figure}
Again, the above equation is equivalent to a set of linear equations for the near horizon components of  $\phi(r)$.
The first four equations are exactly the same as \eqref{scalar_L_1}, \eqref{scalar_L_2}, \eqref{scalar_L_3} and \eqref{scalar_L_4}, however, with $M^{L}_{rs}$ coefficients replaced with $M^{T}_{rs}$ ones.
It turns out that all $M^{T}_{ij}$ coefficients can be generally written as the following:
\begin{equation} 
M^{T}_{ij}(\boldsymbol{\omega},\boldsymbol{k}^2,\boldsymbol{\omega}\boldsymbol{k})=\, i \boldsymbol{\omega} \,a^{T}_{ij}+ \boldsymbol{k}^2 \,b^{T}_{ij}+ d^{T}_{ij}
\end{equation}
with $a^T_{ij}$, $b^T_{ij}$ and $d^T_{ij}$ given in the Table 3 of the Appendix \ref{Tables}.
Repeating the process done in the longitudinal case, we have found the spectrum of the pole-skipping points for the four lowest Matsubara frequencies. The expressions associated with the wavenumbers are complicated as before. Instead of rewriting them, however, we have plotted our results for some specific values of $\nu$ and $b$ in Fig.\ref{chaos_pole_skippin_T_m_02}. In the left panel of the figure we have taken $m=0$ and in the right one $m=2$. As it can be seen, the green (chiral) spectra  have some deviations with respect to the red (non-chiral) ones. The important point, however, is that by entering the chiral effects, the spectra are still symmetric with respect to the vertical axis. This simply means that the information about chiral anomaly, found from the longitudinal poles can not be found from this channel.

\section{Tables}
\label{Tables}
In the following, Table 1 contains the near horizon data associated with the metric and field strength. Table 2 and Table 3 contain the near horizon data associated with the scalar field in the parallel and transverse directions, respectively.
\begin{table}
	\label{table one}
	\centering
	\begin{tabular}[h]{|c|c|}
		\hline
		\hline
		&\\
		Metric functions	&	Evaluated to second order in $b$ and $\nu$ and third order in $\delta=r-r_h$ \\
		& \\
		\hline
		\hline
		$f(r)$&	$4 \pi    T\, (r-r_h)$\\
		&$+ \frac{1}{2}\left(-4+\frac{56 \nu ^2}{3}+b^2 \left(\frac{8}{9} \nu ^2 \left(42 \kappa ^2 (\log (2)-1)-17\right)+\frac{10}{3}\right)\right)(r-r_h)^2$\\
		&	$\,\,+ \frac{1}{6}\left(\frac{24}{\pi  T}-\frac{144 \nu ^2}{\pi  T}+b^2 \left(\frac{8 \nu ^2 \left(\kappa ^2 (363-324 \log (2))+155\right)}{9 \pi  T}-\frac{64}{3 \pi  T}\right)\right)(r-r_h)^3$\\
		\hline             
		$W_{T}(r)$&	$\log (\pi  T)+\frac{2 \nu ^2}{3}+\frac{1}{144} \left(24+\pi ^2\right) b^2$\\
		&	$\,\,\,+  \left(\frac{1}{\pi  T}-\frac{2 \nu ^2}{3 \pi  T}+b^2 \left(\frac{2 \nu ^2 \left(5-6 \kappa ^2 (\log (2)-1)\right)}{9 \pi  T}-\frac{1}{3 \pi  T}\right)\right)(r-r_h)$\\
		&	$\,\,\,\,\,\,\,\,\,+ \frac{1}{2}\left(-\frac{1}{\pi ^2 T^2}+\frac{4 \nu ^2}{3 \pi ^2 T^2}+b^2 \left(\frac{\nu ^2 \left(3 \kappa ^2 (\log (256)-9)-26\right)}{9 \pi ^2 T^2}+\frac{5}{6 \pi ^2 T^2}\right)\right)(r-r_h)^2$\\
		&	$\,\,\,+ \frac{1}{6}\left(\frac{2}{\pi ^3 T^3}-\frac{4 \nu ^2}{\pi ^3 T^3}+b^2 \left(\frac{\nu ^2 \left(\kappa ^2 (91-72 \log (2))+83\right)}{9 \pi ^3 T^3}-\frac{25}{9 \pi ^3 T^3}\right)\right)(r-r_h)^3$\\
		\hline
		$W_{L}(r)$&	$\log (\pi  T)+\frac{2 \nu ^2}{3}+\left(\frac{1}{6}-\frac{\pi ^2}{72}\right) b^2$\\
		&	$+  \left(\frac{1}{\pi  T}-\frac{2 \nu ^2}{3 \pi  T}+b^2 \left(\frac{1}{6 \pi  T}-\frac{2 \nu ^2 \left(\kappa ^2 (\log (64)-6)+1\right)}{9 \pi  T}\right)\right)(r-r_h)$\\
		&	$\,\,\,\,\,+\frac{1}{2} \left(-\frac{1}{\pi ^2 T^2}+\frac{4 \nu ^2}{3 \pi ^2 T^2}+b^2 \left(\frac{2 \nu ^2 \left(6 \kappa ^2 (\log (4)-3)+5\right)}{9 \pi ^2 T^2}-\frac{2}{3 \pi ^2 T^2}\right)\right)(r-r_h)^2$\\
		&	$\,\,\,\,\,\,\,+\frac{1}{6} \left(\frac{2}{\pi ^3 T^3}-\frac{4 \nu ^2}{\pi ^3 T^3}+b^2 \left(\frac{\nu ^2 \left(\kappa ^2 (178-72 \log (2))-22\right)}{9 \pi ^3 T^3}+\frac{23}{9 \pi ^3 T^3}\right)\right)(r-r_h)^3$\\
		\hline
		$E(r)$&	$2 \nu+ b^2 \nu  \left(2 \kappa ^2 (\log (2)-1)-\frac{1}{3}\right)$\\
		&$ + \left(\frac{b^2 \nu  \left(2-2 \kappa ^2 (\log (8)-4)\right)}{\pi  T}-\frac{6 \nu }{\pi  T}\right)(r-r_h)$\\
		&	$\,\,+ \frac{1}{2}\left(\frac{4 b^2 \nu  \left(2 \kappa ^2 (\log (8)-5)-3\right)}{\pi ^2 T^2}+\frac{24 \nu }{\pi ^2 T^2}\right)(r-r_h)^2$\\
		&	$\,\,+ \frac{1}{6}\left(\frac{b^2 \nu  \left(\kappa ^2 (238-120 \log (2))+80\right)}{\pi ^3 T^3}-\frac{120 \nu }{\pi ^3 T^3}\right)(r-r_h)^3$\\
		\hline
		&\\
		$C(r)$&	$\frac{4 b     \nu ^2 \kappa(\log (4)-1)}{\pi  T}(r-r_h)+\frac{2 b    \nu ^2 \kappa(3-10 \log (2))}{\pi ^2 T^2}(r-r_h)^2+\frac{4 b    \nu ^2 \kappa(10\log (2)-1)}{\pi ^3 T^3}(r-r_h)^3$\\
		&\\
		\hline
		&\\
		$P(r)$ & $-b   \nu \kappa+\frac{2 b     \nu \kappa}{\pi  T}(r-r_h)-\frac{5 b    \nu \kappa}{2\pi ^2 T^2}(r-r_h)^2+\frac{5 b \nu  \kappa  }{2 mj\pi ^3 T^3}(r-r_h)^3$\\
		&\\
		\hline
		\hline
	\end{tabular}
	\label{}
	\caption{The near the horizon expanded metric functions \eqref{metric} and field strength \eqref{field_strenght}, given to second order in $\nu$ and $b$.}
\end{table}
\begin{table}
	\label{table_}
	\centering
	\begin{tabular}[h]{|c|c|}
		\hline
		\hline
		Coefficients	&	Evaluated to second order in $b$ and $\nu$ \\
		$a^{L}_{ij}$, $b^{L}_{ij}$, $c^{L}_{ij}$ and $d^{L}_{ij}$	& \\
		\hline
		\hline
		$a^{L}_{11}$&	$	-3 +2  \nu ^2+b^2 \left(2  \nu ^2 \left(\kappa ^2 (\log (4)-2)-1\right)+\frac{1}{2}\right)$\\
		$b^{L}_{11}$&	$	-2+\frac{8 \nu ^2}{3}+b^2 \left(\frac{2}{27} \left(\pi ^2-12\right) \nu ^2+\frac{2}{3}-\frac{\pi ^2}{18}\right)$\\
		$c^{L}_{11}$&	$	4 b   \nu ^2\,\kappa\, (\log (4)-1)$\\
		$d^{L}_{11}$&	$	-(m^2/2)$\\
		\hline             
		$a^{L}_{21}$&	$-\frac{9 }{2 \pi  T}+\frac{6  \nu ^2}{\pi  T}+b^2 \left(\frac{\nu ^2 \left(3 \kappa ^2 (\log (4096)-11)-1\right)}{3 \pi  T}-\frac{1}{2 \pi  T}\right)$\\
		$b^{L}_{21}$&	$-\frac{2}{\pi  T}+\frac{4 \nu ^2}{\pi  T}+b^2 \left(\frac{\nu ^2 \left(4 \left(6 \kappa ^2 (\log (2)-1)-1\right)+\pi ^2\right)}{9 \pi  T}-\frac{\pi ^2-6}{18 \pi  T}\right)$\\
		$c^{L}_{21}$&	$	\frac{b   \nu ^2 \,\kappa\,(8\log (2)-6)}{\pi  T}$\\
		$d^{L}_{21}$&	$-\frac{m^2}{\pi  T}+\frac{2 m^2 \nu ^2}{3 \pi  T}+b^2 \left(\frac{2 m^2 \nu ^2 \left(6 \kappa ^2 (\log (2)-1)+1\right)}{9 \pi  T}-\frac{m^2}{6 \pi  T}\right)$\\
		\hline
		$a^{L}_{22}$&	$-\frac{11 }{2}+\frac{11 \nu ^2}{3}+b^2 \left(\frac{1}{9} \nu ^2 \left(66 \kappa ^2 (\log (2)-1)-1\right)-\frac{5 }{12}\right)$\\
		$b^{L}_{22}$&	$-1+\frac{4 \nu ^2}{3}+b^2 \left(\frac{1}{27} \left(\pi ^2-12\right) \nu ^2+\frac{1}{3}-\frac{\pi ^2}{36}\right)$\\
		$c^{L}_{22}$&	$	2 b   \nu ^2\,\kappa\, (\log (4)-1)$\\
		$d^{L}_{22}$&	$6-\frac{m^2}{4}+b^2 \left(1-\frac{8 \nu ^2}{3}\right)$\\
		\hline
		$a^{L}_{31}$&	$-\frac{3 }{\pi ^2 T^2}+\frac{6 \nu ^2}{\pi ^2 T^2}+b^2 \left(\frac{4 \nu ^2 \left(\kappa ^2 (9\log (2)-7)+1\right)}{3 \pi ^2 T^2}-\frac{1}{2 \pi ^2 T^2}\right)$\\
		$b^{L}_{31}$&	$-\frac{2}{3 \pi ^2 T^2}+\frac{16 \nu ^2}{9 \pi ^2 T^2}+b^2 \left(\frac{4 \nu ^2 \left(9 \left(2 \kappa ^2 (\log (4)-1)-1\right)+\pi ^2\right)}{81 \pi ^2 T^2}-\frac{\pi ^2-12}{54 \pi ^2 T^2}\right)$\\
		$c^{L}_{31}$&	$	\frac{2 b   \nu ^2 \,\kappa\,(\log (16)-3)}{3 \pi ^2 T^2}$\\
		$d^{L}_{31}$&	$-\frac{m^2}{\pi ^2 T^2}+\frac{4 m^2 \nu ^2}{3 \pi ^2 T^2}+b^2 \left(\frac{2 m^2 \nu ^2 \left(6 \kappa ^2 (\log (64)-5)+7\right)}{27 \pi ^2 T^2}-\frac{2 m^2}{9 \pi ^2 T^2}\right)$\\
		\hline
		$a^{L}_{32}$&	$-\frac{7 }{\pi  T}+\frac{28  \nu ^2}{3 \pi  T}+b^2 \left(\frac{2 \nu ^2 \left(3 \kappa ^2 (84 \log (2)-73)+25\right)}{27 \pi  T}-\frac{11 }{9 \pi  T}\right)$\\
		$b^{L}_{32}$&	$-\frac{4}{3 \pi  T}+\frac{8 \nu ^2}{3 \pi  T}+b^2 \left(\frac{2 \nu ^2 \left(4 \left(6 \kappa ^2 (\log (2)-1)-1\right)+\pi ^2\right)}{27 \pi  T}-\frac{\pi ^2-6}{27 \pi  T}\right)$\\
		$c^{L}_{32}$&	$\frac{4 b   \nu ^2 \,\kappa\,(\log (16)-3)}{3 \pi  T}$\\
		$d^{L}_{32}$&	$\frac{25-2 m^2}{3 \pi  T}+\frac{4 \left(m^2-13\right) \nu ^2}{9 \pi  T}+b^2 \left(\frac{2 \nu ^2 \left(3 \kappa ^2 (51-52 \log (2))+2 m^2 \left(6 \kappa ^2 (\log (2)-1)+1\right)-65\right)}{27 \pi  T}-\frac{m^2-23}{9 \pi  T}\right)$\\
		\hline
		$a^{L}_{33}$&	$-\frac{19 }{3}+\frac{38  \nu ^2}{9}+b^2 \left(\frac{2}{27} \nu ^2 \left(114 \kappa ^2 (\log (2)-1)+7\right)-\frac{13 }{18}\right)$\\
		$b^{L}_{33}$&	$-\frac{2}{3}+\frac{8 \nu ^2}{9}+b^2 \left(\frac{2}{81} \left(\pi ^2-12\right) \nu ^2+\frac{2}{9}-\frac{\pi ^2}{54}\right)$\\
		$c^{L}_{33}$&	$\frac{4}{3} b   \nu ^2 \,\kappa\,(\log (4)-1)$\\
		$d^{L}_{33}$&	$\frac{38}{3}-\frac{m^2}{6}-\frac{4 \nu ^2}{9}+b^2 \left(\frac{25}{9}-\frac{4}{27} \nu ^2 \left(6 \kappa ^2 (\log (2)-1)+49\right)\right)$\\
		\hline
		\hline
	\end{tabular}
	\label{}
	\caption{Longitudinal dynamics of scalar field.}
\end{table}
\begin{table}
	\label{table}
	\centering
	\begin{tabular}[h]{|c|c|}
		\hline
		\hline
		&\\
		Coefficients	&	Evaluated to second order in $b$ and $\nu$ \\
		$a^{T}_{ij}$, $b^{T}_{ij}$ and $d^{T}_{ij}$	& \\
		\hline
		\hline
		$a^{T}_{11}$&	$-3 +2  \nu ^2+b^2 \left(2 \nu ^2 \left(\kappa ^2 (\log (4)-2)-1\right)+\frac{i}{2}\right)$\\
		$b^{T}_{11}$&	$-2+\frac{8 \nu ^2}{3}+b^2 \left(\left(-\frac{8}{9}-\frac{\pi ^2}{27}\right) \nu ^2+\frac{1}{36} \left(24+\pi ^2\right)\right)$\\
		$d^{T}_{11}$&	$	-(m^2/2)$\\
		\hline             
		$a^{T}_{21}$&	$-\frac{9 }{2 \pi  T}+\frac{6  \nu ^2}{\pi  T}+b^2 \left(\frac{\nu ^2 \left(3 \kappa ^2 (\log (4096)-11)-1\right)}{3 \pi  T}-\frac{1}{2 \pi  T}\right)$\\
		$b^{T}_{21}$&	$-\frac{2}{\pi  T}+\frac{4 \nu ^2}{\pi  T}+b^2 \left(\frac{\pi ^2-24}{36 \pi  T}-\frac{\nu ^2 \left(\pi ^2-16 \left(\kappa ^2 (\log (8)-3)+4\right)\right)}{18 \pi  T}\right)$\\
		$d^{T}_{21}$&	$-\frac{m^2}{\pi  T}+\frac{2 m^2 \nu ^2}{3 \pi  T}+b^2 \left(\frac{m^2 \nu ^2 \left(3 \kappa ^2 (\log (16)-4)+2\right)}{9 \pi  T}-\frac{m^2}{6 \pi  T}\right)$\\
		\hline
		$a^{T}_{22}$&	$-\frac{11 }{2}+\frac{11 \nu ^2}{3}+b^2 \left(\frac{1}{9} \nu ^2 \left(66 \kappa ^2 (\log (2)-1)-1\right)-\frac{5}{12}\right)$\\
		$b^{T}_{22}$&	$-1+\frac{4 \nu ^2}{3}+b^2 \left(\left(-\frac{4}{9}-\frac{\pi ^2}{54}\right) \nu ^2+\frac{1}{72} \left(24+\pi ^2\right)\right)$\\
		$d^{T}_{22}$&	$6-\frac{m^2}{4}+b^2 \left(1-\frac{8 \nu ^2}{3}\right)$\\
		\hline
		$a^{T}_{31}$&	$-\frac{3 }{\pi ^2 T^2}+\frac{6\nu ^2}{\pi ^2 T^2}+b^2 \left(\frac{4\nu ^2 \left(\kappa ^2 (\log (512)-7)+1\right)}{3 \pi ^2 T^2}-\frac{1}{2 \pi ^2 T^2}\right)$\\
		$b^{T}_{31}$&	$-\frac{2}{3 \pi ^2 T^2}+\frac{16 \nu ^2}{9 \pi ^2 T^2}+b^2 \left(\frac{\pi ^2-12}{108 \pi ^2 T^2}-\frac{2 \nu ^2 \left(9 \left(\kappa ^2 (1-8 \log (2))-8\right)+\pi ^2\right)}{81 \pi ^2 T^2}\right)$\\
		$d^{T}_{31}$&	$-\frac{m^2}{\pi ^2 T^2}+\frac{4 m^2 \nu ^2}{3 \pi ^2 T^2}+b^2 \left(\frac{2 m^2 \nu ^2 \left(6 \kappa ^2 (\log (64)-5)+7\right)}{27 \pi ^2 T^2}-\frac{2 m^2}{9 \pi ^2 T^2}\right)$\\
		\hline
		$a^{T}_{32}$&	$-\frac{7 }{\pi  T}+\frac{28 \nu ^2}{3 \pi  T}+b^2 \left(\frac{2\nu ^2 \left(3 \kappa ^2 (84 \log (2)-73)+25\right)}{27 \pi  T}-\frac{11}{9 \pi  T}\right)$\\
		$b^{T}_{32}$&	$-\frac{4}{3 \pi  T}+\frac{8 \nu ^2}{3 \pi  T}+b^2 \left(\frac{\pi ^2-24}{54 \pi  T}-\frac{\nu ^2 \left(\pi ^2-16 \left(\kappa ^2 (\log (8)-3)+4\right)\right)}{27 \pi  T}\right)$\\
		$d^{T}_{32}$&	$\frac{25-2 m^2}{3 \pi  T}+\frac{4 \left(m^2-13\right) \nu ^2}{9 \pi  T}+b^2 \left(\frac{2 \nu ^2 \left(3 \kappa ^2 (51-52 \log (2))+2 m^2 \left(6 \kappa ^2 (\log (2)-1)+1\right)-65\right)}{27 \pi  T}-\frac{m^2-23}{9 \pi  T}\right)$\\
		\hline
		$a^{T}_{33}$&	$-\frac{19 }{3}+\frac{38  \nu ^2}{9}+b^2 \left(\frac{2}{27}  \nu ^2 \left(114 \kappa ^2 (\log (2)-1)+7\right)-\frac{13 }{18}\right)$\\
		$b^{T}_{33}$&	$-\frac{2}{3}+\frac{8 \nu ^2}{9}+b^2 \left(\frac{1}{81} \left(-24-\pi ^2\right) \nu ^2+\frac{1}{108} \left(24+\pi ^2\right)\right)$\\
		$d^{T}_{33}$&	$\frac{38}{3}-\frac{m^2}{6}-\frac{4 \nu ^2}{9}+b^2 \left(\frac{25}{9}-\frac{4}{27} \nu ^2 \left(6 \kappa ^2 (\log (2)-1)+49\right)\right)$\\
		\hline
		\hline
	\end{tabular}
	\label{}
	\caption{Transverse dynamics of scalar field.}
\end{table}
\newpage
\bibliographystyle{utphys}
\providecommand{\href}[2]{#2}\begingroup\raggedright\endgroup

\end{document}